\documentclass[10pt,a4paper]{iopart}
\usepackage{iopams}  
\usepackage{graphicx,psfrag,bbm,latexsym,color,dcolumn,bm,dsfont,bbm,color,
mathrsfs,bbold,latexsym,amsfonts,amssymb}

\def\arctanh{\mathop{\rm arctanh}\nolimits}

\newcommand{\mediau}[1]{\int d\mu~ #1}
\newcommand{\media}[1]{\int d\mu_b~ #1}

\newcommand{\mediaT}[1]{\left\langle #1 \right\rangle}
\newcommand{\mediaTc}[1]{\left\langle #1 \right\rangle^{(c)}}

\begin{document}


\title[Ising spin glass models versus Ising models I]
{Ising spin glass models versus Ising models: \\
an effective mapping at high temperature I. \\
General result}

\author{Massimo Ostilli$^{1,2}$}
\address{$^1$\ Departamento de Fisica,
Universidade de Aveiro,
Campus Universitario de Santiago 3810-193, Aveiro, Portugal.}
\address{$^2$\ Center for Statistical Mechanics and Complexity, Istituto 
Nazionale per la Fisica della Materia, 
Unit\`a di Roma 1, Roma 00185, Italy.}

\ead{massimo.ostilli@roma1.infn.it}

\date{\today}

\begin{abstract}
We show that, above the critical temperature, 
if the dimension $D$ of a given Ising spin glass model 
is sufficiently high, its free energy can
be effectively expressed through the free energy of  
a related Ising model. When, in a large sense, $D=\infty$, 
in the paramagnetic phase and on its boundary the mapping is exact.
In this limit the method provides a general and simple rule to obtain
exactly the upper phase boundaries. 
We provide even simple effective rules to find crossover surfaces 
and correlation functions. 
We apply the mapping to several spin glass models.
\end{abstract}

\pacs{05.20.-y, 75.10.Nr, 05.70.Fh, 64.60.-i, 64.70.-i}

\maketitle

\section{Introduction}

The spin glass model represents one of the most difficult
challenges in physics (see Ref. \cite{Parisi} and referred articles). 
When compared to a non random model,
the spin glass model turns out to be hugely more difficult to solve.
In fact, to solve a spin glass model, one should be able to study, at least in
principle, either a non random model, but with non uniform frozen couplings, 
or, as appears evident in the framework of the replica approach, 
a system of $n$ coupled uniform non random
models, extrapolating some suitable derivative 
in the limit $n\rightarrow 0$.
From this point of view it seems fairly hard that a spin glass might
be solved simply in terms of a mapping with a non random and uniform model.
Nevertheless, in this paper we show that, 
whereas in the most rich and complex low temperature phases
such a mapping is impossible,
when the dimensionality $D$ of the model tends to infinite, 
in the paramagnetic phase and on its boundary, the given spin
glass model can be easily solved by simply considering a suitable
corresponding non random Ising model: ``the related Ising model''.
More precisely, we shall show that, when in a large sense the dimensionality
$D$ is infinite, as happens, \textit{e.g.}, but not only, 
in mean field models and in generalized tree-like structures 
(allowing also for the presence of loops), once the phase boundary
of the related Ising model is known, it can be immediately used in the spin
glass model to find
exactly the region of the paramagnetic phase ($P$) and its boundaries
with the other phases: ferromagnetic disordered ($F$), spin glass ($SG$), 
or antiferromagnetic disordered ($AF$). 

The derivation of this mapping, ``spin glass'' $\rightarrow$ 
``related Ising model'', will be obtained in the framework of 
a replica approach, which however differs from the standard one \cite{Parisi}. 
As we shall show, the use of replicas in the high temperature expansion
of the free energy leads to a different procedure in which
there is no functional to be extremized and, furthermore,
when $D$ is large, at high temperature a simple combinatorial 
approximation applies which, in the limit $D\rightarrow \infty$, 
becomes exact.
We point out that, unlike the standard replica approach,
in the proof for this mapping we do not need to rely
on any ansatz concerning the choice of finding stationary points.
Therefore our proof, up to question of the analytic
continuation $n$ integer $\rightarrow$ $n$ real, 
is exact \cite{NOTA0}-\cite{Franz}.

Previous uses of the high temperature expansion
to study spin glass models go back to \cite{Palmer} for the
Sherrignton-Kirkpatrick model (SK) \cite{SK}, and to \cite{Singh}
and \cite{Fisher1} for the Ising spin glass models on hypercubic lattices 
with simmetric disorder, where very accurate results were
found for small and high dimensions in \cite{Singh} and \cite{Fisher1},
respectively (we cite also the Ref. \cite{Janke} and the 
Refs. \cite{Potts,Potts1} for the Potts glass model).
However, as will be shown, our use of the high temperature expansion of
the partition function is quite different. 
We do not use this expansion to directly calculate the
averages over the disorder of a specific model, 
we use instead the expansion to find a
general link between the spin glass and a suitable related Ising model.
Once this link has been established, the singularities of the related
Ising model will provide the singularities of the spin glass,  
signaling the phase transitions in the sense of Lee and Yang \cite{Lee}.

The practical potential applications of this mapping are remarkable.
In fact, though the mapping may be used to analyze only the regions at
high temperature, including the upper phase 
boundary and, within a certain approximation, the crossover
surfaces and the correlation functions, 
as will be clear, its generality and simplicity 
make the approach particularly suitable to face those
infinite dimensional spin glass models whose great complexity,
\textit{e.g.}
due to the presence of too many parameters, can make difficult, or avoids
at all, the application of the standard methods for spin glasses
even in the easier paramagnetic phase.
In our approach instead, the model to be solved is the related Ising model
which, as a non random model, turns out to be hugely easier
to solve, analytically and/or numerically. 
In this work (part I) we will show this in 
several non trivial known examples,
whereas in a forthcoming paper (part II) the mapping will be applied 
for studing the Ising spin glass model on general graphs and networks.

The paper is organized as follows. 
In Secs. 2 and 3 we introduce the models and the condition under which
the mapping holds. 
In Sec. 4, after introducing the definition of the related Ising model, 
we present formally the mapping, subdiving the results 
in the subsections 4.1 (same disorder for any bond), 
4.2 (generalization), 4.3 (upper line of the phase diagram).
As a by-product, in the further subsection 4.4, via analytic continuation, 
we, improperly, force the mapping to find crossover surfaces 
and correlation functions.
We stress that this extension of the mapping is not exact,
but provide however a first effective insight about the physics of the model.

The rest of the article until Sec. 11 is devoted to the proof of the mapping.
In Sec. 5 we give some preliminary ideas, whereas in 
Sec. 6 we recall the high temperature expansion for a general Ising 
(even non uniform) model. In Sec. 7, starting form the high temperature
expansion, we carry out formally the average over the disorder to be used
in Sec. 8 for calculating the free energy.
In Sec. 9 we specialize the expansion for a centered measure, whereas
in the following subsection 9.1 the basic approximation in high dimension
will be introduced and used to find the mapping for a hypercube lattice.
In Sec. 10 and in the following subsection we generalize the mapping
to any measure; finally, in Sec. 11 we extend the mapping to any model
whose dimension, in a large sense, turns out to be infinite,
as happens, in particular, but not only,
in tree-like structures and generalized tree-like structures.

The sections 12 and 13 are devoted to some applications in the case
of $D$ finite and infinite, respectively.
We anticipate that our approach takes into account only the leading
term of a $1/D$ expansion, so that 
the applications in the finite dimensional case are basically meant
to show how the method works.

Finally, conclusions and some outlooks are reported in Sec. 14.
The paper is equipped with three appendix.

\section{Models}

\label{model}
Let us consider a $D$ dimensional hypercube of side $L$, 
$\Lambda=\{1,\ldots,L\}^D$, and the set of the bonds $b$
connecting two first neighbors sites
\begin{eqnarray}
\label{Gamma}
\Gamma\equiv \{b=\left(i_b,i_b\right): i_b,j_b \in 
\Lambda,~ i_b~ \mbox{and}~ j_b~ \mbox{first neighbors},~ i_b<j_b\}.
\end{eqnarray} 
We will indicate with $N$ the total number of points, $N=L^D$.
More in general, we shall consider also systems over a graph. 
Given a graph $g$ of $N$ vertices, the set of links will be
defined through the adjacency matrix of the graph, $g_{i,j}=0,1$:
\begin{eqnarray}
\label{Gamma1}
\Gamma\equiv \{b=\left(i_b,i_b\right): i_b,j_b \in g,
~ g_{i_b,j_b}=1,~ i_b<j_b\}.
\end{eqnarray} 
The set of links of the 
fully connected graph will be indicated with $\Gamma_f$:
\begin{eqnarray}
\label{Gammaf}
\Gamma_f\equiv \{b=\left(i_b,i_b\right): i_b,j_b=1,\ldots,N, ~ i_b<j_b\}.
\end{eqnarray} 

The Hamiltonian of the spin glass with two-body interactions can be written as
\begin{eqnarray}
\label{H}
H\left(\{\sigma_i\};\{J_b\};\{h_i\}\right)
\equiv -\sum_{b\in\Gamma} J_b \tilde{\sigma}_b+\sum_{i=1}^N h_i \sigma_i,
\end{eqnarray} 
where 
the $h_i$'s are arbitrary 
external fields, the $J_b$'s are 
quenched couplings, $\sigma_i$ is an Ising variable at the site $i$, and 
$\tilde{\sigma}_b$ stays for the product
of two Ising variables, $\tilde{\sigma}_b=\sigma_{i_{b}}\sigma_{j_{b}}$, with
$i_{b}$ and $j_{b}$ such that $b=\left(i_b,j_b\right)$. 

The free energy $F$ is defined by
\begin{eqnarray}
\label{logZ}
-\beta F\equiv \int d\mathcal{P}\left(\{J_b\}\right)
\log\left(Z\left(\{J_b\};\{h_i\}\right)\right),
\end{eqnarray} 
where $Z\left(\{J_b\};\{h_i\}\right)$ 
is the partition function of the quenched system
\begin{eqnarray}
\label{Z}
Z\left(\{J_b\};\{h_i\}\right)= \sum_{\{\sigma_b\}}e^{-\beta 
H\left(\{\sigma_i\};\{J_b\};\{h_i\}\right)}, 
\end{eqnarray} 
and $d\mathcal{P}\left(\{J_b\}\right)$ 
is a product measure over all the possible bonds $b$ given 
in terms of normalized measures $d\mu_b\geq 0$ 
(we are considering a general measure $d\mu_b$ 
allowing also for a possible dependence on the bonds) 
\begin{eqnarray}
\label{dP}
d\mathcal{P}\left(\{J_b\}\right)\equiv \prod_{b\in\Gamma_f} 
d\mu_b\left( J_b \right),
\quad \int d\mu_b\left( J_b \right) =1.
\end{eqnarray}

Among the measures most considered in literature, we cite in particular the
Gaussian measure 
\begin{eqnarray}
\label{dmugauss0}
\frac{d\mu\left(J_b\right)}{dJ_b} = \frac{1}{\sqrt{2\pi \tilde{J}^2}}
\exp\left(-J_b^2/\tilde{J}^2\right),
\end{eqnarray} 
and the ``plus-minus'' measure
\begin{eqnarray}
\label{plusminus}  
\frac{d\mu(J_b)}{dJ_b} = \frac{1}{2}\delta(J-J_b)+\frac{1}{2}\delta(J+J_b),
\end{eqnarray}
where $\tilde{J}^2$ and $J$ represent disorder parameters.
We will take the Boltzmann constant $K_B=1$. 
A generic inverse critical temperature of the spin glass model, if any, 
will be indicated 
with $\beta_c$; sometimes we will use the symbol $<A>$ 
for the quenched thermal average of the quantity $A$ (fixed 
values of the couplings $\{J_b\}$);
finally the density free energy in the thermodynamic 
limit will be indicated with $f=f(\beta)$
\begin{eqnarray}
\label{f}
f(\beta)\equiv \lim_{N\rightarrow \infty}F(\beta)/N.
\end{eqnarray} 

\section{Dimensionality - Condition for the mapping}
Our mapping will be first derived in a $D$ dimensional
hypercubic lattice by applying an approximation becoming
exact in the limit $D\to\infty$. 
Successively, we will consider more general structures, whose
dimension, in a large sense, goes to infinite as well.
For a hypercube lattice the dimension $D$ is related to the
number of first neighbors of a vertex, $2D$, so that, in this case,
and in others in which the numbers of first neighbors is proportional to
$D$, $D\to\infty$ if the number of first neighbors goes to infinite
For these cases, we shall say that the set of links $\Gamma$ is
\textit{infinite dimensional in a strict sense}.
For other structures, a usual definition of dimensionality is 
\begin{eqnarray}
\label{D}
D\equiv \lim_{l\to\infty}\frac{\log\left(N(l)\right)}{\log\left(l\right)},
\end{eqnarray} 
where $N(l)=1+m_1+\dots +m_l$ represents the total number of vertices
within $l$ steps of an arbitrary fixed root vertex 0, 
$m_k$ being the number of vertices at distance $k$ from the root. 
We will not make use of this definition for the dimension, but here
we just observe what follows.
When applied to a hypercube lattice, Eq. (\ref{D}) returns the true
dimension of the lattice $D$, whereas in the case of a tree with
an average degree greater than 1, Eq. (\ref{D}) gives $D=\infty$. 
As an heuristic argument we can say that, given a structure $\Gamma$,
the hypercube lattice most similar to $\Gamma$ must have a dimension
$D$ given by Eq. (\ref{D}); therefore, if, for $\Gamma$, 
$D$ of Eq. (\ref{D}) goes to infinite, the mapping must hold.
However, we will not need to consider this heuristic argument.
In Sec. 11, we will work out directly the derivation of the mapping
for tree-like structures, little differences being involved
with respect to the hypercubic lattice. Furthermore, we shall now
introduce the most general condition under which the mapping is exact,
stressing that the tree-like structure is not necessary.
 
Let us define a path as a succession of first neighbor vertices
connected by different bonds of $\Gamma$. A bond has length 1. 
The paths can be open or closed (loops). 
As will be clear later, depending on $\Gamma$ and near the limit
$\beta\to\beta_{c}^{-}$, the paths
giving a finite contribution to the partition function $Z$ are 
infinitely long paths which can be:
closed, as for hypercubic lattices; open, as for tree-like structures;
and both closed and open, as for generalized tree-like structures.
We say that $\Gamma$ has a generalized tree-like structure if for any vertex 
there is at most a finite number of loops.

\begin{figure}[t]
\centering
\includegraphics[width=0.5\columnwidth,clip]{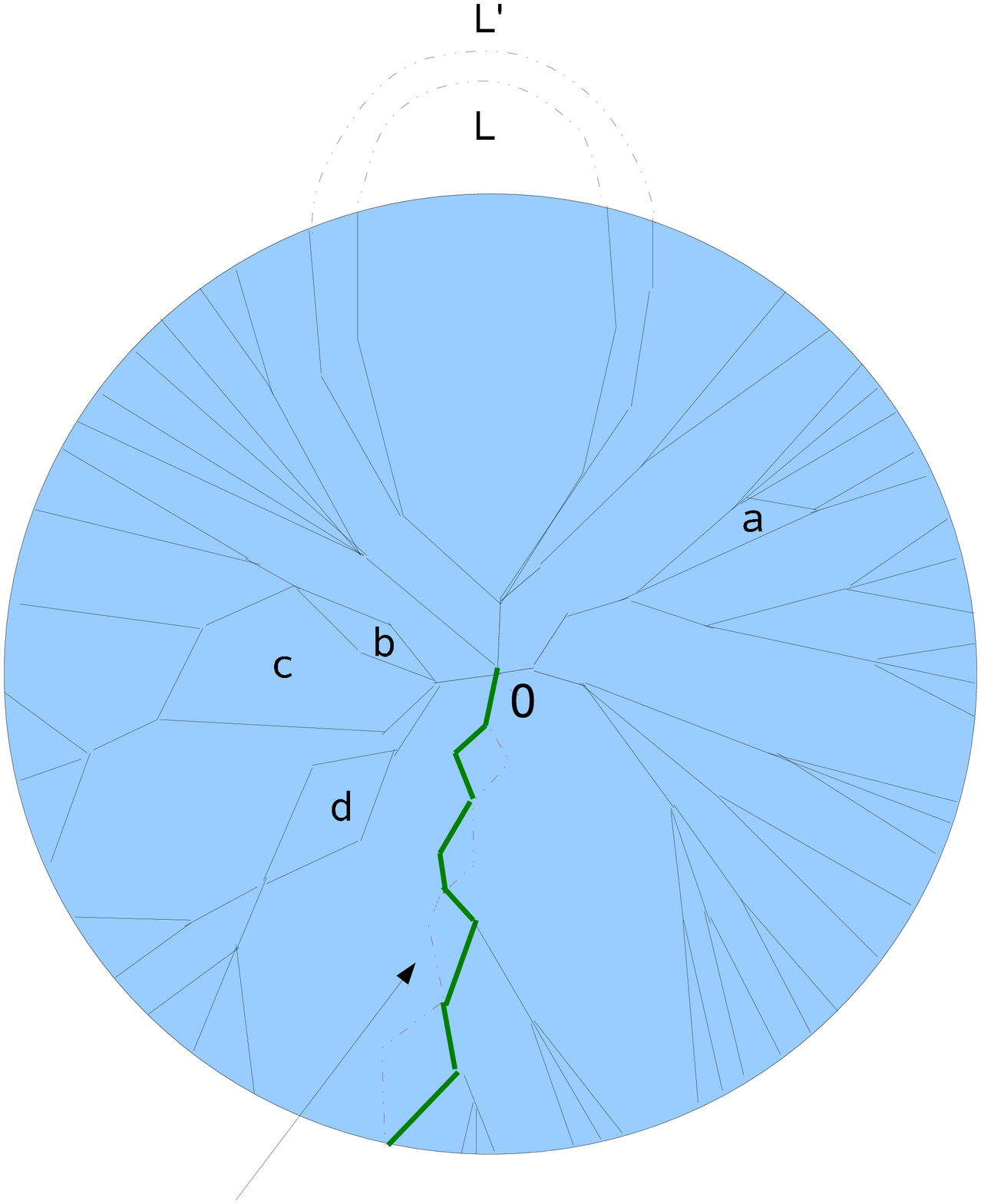}
\caption{ 
Example to clarify the condition given in Sec. 3. 
From the root vertex 0 depart several infinitely long paths.
For simplicity, in the figure it is understood that, except for
the chain indicated by an arrow, besides the loops
a, b, c, d, L and L', there are no other loops connected to the root 0.
It is also understood that some branching of the paths keeps on also
out of the circle.
From each of the loops a, b, c, and d, and from the other branches
without loops, pass an infinite number
of infinitely long paths, but the probability that two of them,
randomly chosen, overlap for an infinite number of bonds, goes to 0.
Through the chain of loops (infinite chain) pass also an infinite
number of infinitely long paths, but in this case, the probability
that, two of them randomly chosen, overlap for an infinite 
number of bonds is not 0.
A way to generate, from the above graph, a graph infinite dimensional
in the large sense, consists in deleting, \textit{e.g.}, 
all the infinite bonds of the chain indicated with dashed segments.
The loops L and L' are two example of infinite closed paths.
Note that, since they share only a finite number of bonds each other,
their presence do not alter the condition of Sec. 3.} 
\label{ball}
\end{figure}

%
%
What will emerge in subsection 9.1 and Sec. 11, is that 
\begin{center}
\textit{when, in the
thermodynamic limit, for the given set $\Gamma$, 
the total number of infinitely long paths per vertex 
goes to infinite and, choosing randomly two of them,
the probability that they overlap each other 
for an infinite number of bonds goes to zero,
in the limit $\beta\to\beta_{c}^{-}$, the mapping becomes exact.}
\end{center}
We shall say that such a $\Gamma$ is 
\textit{infinite dimensional in the large sense}.
It is immediate to verify that in the case of tree-like structures
the above condition is trivially satisfied and similarly
for generalized tree-like structures.
In general, given $\Gamma$, the absence of loops, or the presence
of a finite number of loops per vertex,
turns out to be a sufficient condition for 
satisfying the above requirement.
Note however that this is not a necessary condition.
Loops can also be massively present; what is necessary is only
that loops overlap, either each other or with other paths, only partially
(see Fig. \ref{ball}).
Of course such a behavior does not 
happen in a hypercube lattice of finite dimension $D$. 
It is in fact easy to see that, in this case,
the total number of paths of length $l$ per vertex 
grows as $c(l)\sim\exp(\mu_D l)$, 
with $\mu_D$ a suitable constant \cite{Fisher}, 
whereas the maximum number of paths which
after $l$ steps do not share any other bond
grows only as $c_{i}(l)\sim l^D$. 
Therefore, if $D$ is finite, in the limit $l\to\infty$, we
see that the condition cannot be satisfied.

%

\section{The mapping}
Given a spin glass model trough Eqs. (\ref{Gamma1}-\ref{dP}), 
we define, on the same set of links $\Gamma$, its \textit{related Ising model} 
trough the following Ising Hamiltonian
\begin{eqnarray}
\label{HI}
H_I\left(\{\sigma_i\};\{J_b\};\{h_i\}\right)
\equiv -\sum_{b\in\Gamma} J_b^{(I)} \tilde{\sigma}_b+\sum_{i=1}^N h_i \sigma_i
\end{eqnarray} 
where the Ising couplings $J_b^{(I)}$'s have 
non random values such that $~\forall ~b,b'\in \Gamma$
\begin{eqnarray}
\label{JI}
J_{b'}^{(I)}&=&J_b^{(I)} \quad \mathrm{if} \quad 
d\mu_{b'}\equiv d\mu_{b}, \\
\label{JIb}
J_b^{(I)}&\neq & 0 \quad \mathrm{if} \int d\mu_b(J_b)J_b\neq 0 \quad
\mathrm{or} \quad \int d\mu_b(J_b)J_b^2>0. 
\end{eqnarray}
In the following a suffix $I$ over quantities such as $H_{I}$,
$F_{I}$, $f_{I}$, etc\ldots, or $J_b^{(I)}$, $\beta_c^{(I)}$, etc\ldots,
will be referred to the related Ising system with Hamiltonian (\ref{HI}).
We can always split the free energy of the quenched model
as follows
\begin{eqnarray}
\label{0logZ2}
-\beta F=\sum_{b\in\Gamma} \int d\mu_{b} \log\left(2\cosh(\beta J_b)\right)
 + \sum_{i=1}^N \log\left(2\cosh(\beta h_i)\right)+ \phi,
\end{eqnarray}
$\phi$ being the high temperature part of the free energy, and similarly
for the related Ising model.
Let $\varphi$ be the density of $\phi$ in the thermodynamic limit
\begin{eqnarray}
\label{varphi}
\varphi \equiv \lim_{N\rightarrow \infty}\phi/N,
\end{eqnarray} 
and similarly for $\varphi_{I}$, the high temperature part
of the free energy density of the related Ising model defined through
Eqs. (\ref{HI}-\ref{JIb}). 
As is known, the high temperature part $\varphi$ ($\varphi_I$) 
can be expressed in terms of the
quantities $z_b=\tanh(\beta J_b)$ 
($z_b^{(I)}=\tanh(\beta J_b^{(I)})$) and $z_i=\tanh(\beta h_i)$, 
\textit{i.e.}, 
the adimensional parameters of the Ising high temperature expansion:
\begin{eqnarray}
\label{varphi0}
\varphi &=& \varphi
\left(\{\tanh(\beta J_b)\};\{\tanh(\beta h_i)\}\right),\\
\label{varphi1}
\varphi_I&=& \varphi_I
\left(\{\tanh(\beta J_b^{(I)})\};\{\tanh(\beta h_i)\}\right),
\end{eqnarray} 

Hereafter, if not explicitly said, we set $h_i\equiv 0$. 
By varying in the Ising Hamiltonian (\ref{HI}) the couplings $J_b^{(I)}$ 
with the constrains (\ref{JI}-\ref{JIb}), 
we explore the function $\varphi_{I}$ and
the non analytic points of $\varphi_{I}$, if any, will signal
some phase transition.
The critical behavior of $\varphi_{I}$ will
be characterized by an equation of the type $G_I(\{z_b^{(I)}\})=0$,
whose solution is provided only by universal, \textit{i.e.}, not depending on 
the $\{J_b^{(I)}\}$, quantities. 
A point on the critical surface $\Sigma_I$, solution of the equation 
$G_I=0$, will be indicated with $\{w_{b}^{(I)}\}$. 
The surface $\Sigma_I$ represents the boundary of some domain 
$\mathscr{D}_I$ inside which the high temperature
expansion providing $\varphi_{I}$ converges. 
As we shall discuss later, the domain $\mathscr{D}_I$ is a convex set 
so that, if $|z_b^{(I)}|=|\tanh(\beta J_b^{(I)})|<w_b^{(I)}$ for any $b$, then 
$\{z_b^{(I)}\}\in \mathscr{D}_I$. 

Our main result concerns $\varphi$ and its singularities, \textit{i.e.}
the phase boundaries of the spin glass model.
We find convenient to separate these results in 
the cases of homogeneous and inhomogeneous measures.

\subsection{Case of an homogeneous measure (same disorder for any bond)}
Let be $d\mu_b\equiv d\mu$ for any bond $b$ of $\Gamma$,
so that, according to Eqs. (\ref{JI}) and (\ref{JIb}), 
the related Ising model corresponds to a homogeneous Ising model
having a single coupling $J_b^{(I)}\equiv J^{(I)}$
and its critical behavior will be characterized by, at most, two points
$w_{F}^{(I)}>0$ and $w_{AF}^{(I)}<0$, if any. 

First, let us consider a system having a number of first neighbors per vertex
proportional to $D$.
Let be $\beta_c^{(SG)}$ and $\beta_c^{(F/AF)}$, respectively, 
the solutions of the equations, if any
\begin{eqnarray}
\label{mapp0}
\mediau{\tanh^2(\beta_c^{(SG)} J_b)}&=&w_{F}^{(I)}, 
\qquad D>2,\\
\label{mapp01}
\mediau{\tanh(\beta_c^{(F/AF)} J_b)}&=&w_{F/AF}^{(I)},
\qquad D>1,
\end{eqnarray} 
where $F$ or $AF$, in the l.h.s and r.h.s. of Eq. (\ref{mapp01}), 
are in correspondence.  
We will show that, asymptotically, at high dimensions $D$, 
the critical inverse temperature of the spin glass model, $\beta_c$,
is given by
\begin{eqnarray}
\label{mapp}
\beta_c=\mathrm{min}\{\beta_c^{(SG)},\beta_c^{(F/AF)}\};
\end{eqnarray} 
and in the paramagnetic phase the following mapping holds 
\begin{eqnarray}
\label{mapp1}
\left|\frac{\varphi-\varphi_{eff}}{\varphi}\right|=
O\left(\frac{g(\beta^{-1}-\beta_c^{-1})}{D}\right), 
\quad \forall \beta\leq \beta_c,
\end{eqnarray} 
where, $g(x)$ is some bounded continuous function of order 1 such that 
$g(x)\rightarrow 0$ for $x\rightarrow \infty$;
and $\varphi_{eff}$, is given by
\begin{eqnarray}
\label{mapp2}
\varphi_{eff}=\frac{1}{l}
\varphi_{I}\left(\mediau{\tanh^{l}(\beta J_b)}\right), 
\qquad D>2^{l-1},
\end{eqnarray}
where
\begin{eqnarray}\fl
\label{mapp3}
l=\left\{
\begin{array}{l}
2, \quad \mathrm{if}\quad
|\varphi_{I}\left(\mediau{\tanh^{2}(\beta J_b)}|\right) \geq 
2|\varphi_{I}\left(\mediau{\tanh(\beta J_b)}\right)|,\\
1, \quad \mathrm{if}\quad
|\varphi_{I}\left(\mediau{\tanh^{2}(\beta J_b)}|\right) < 
2|\varphi_{I}\left(\mediau{\tanh(\beta J_b)}\right)|.
\end{array}
\right.
\end{eqnarray} 
If the system $\Gamma$ is infinite dimensional 
only in the large sense (Sec. 3),
in the above equations $D$ can be settled as infinite and, 
in the limit $\beta\to\beta_c^{-}$, 
Eq. (\ref{mapp1}) holds exactly.

\subsection{Generalization}
The generalization of the above mapping to the case of an arbitrary 
(not homogeneous) measure $d\mu_b$, useful for example for anisotropic
models, in which we have to consider even a given number of bond dependencies,
follows straightforward. 
In this case, the related Ising model is defined by a set of,
typically few, independent couplings $\{J_b^{(I)}\}$, 
trough Eqs. (\ref{JI}-\ref{JIb})
and its critical behavior will be fully characterized 
by the points of $\Sigma_I$, solution of the equation $G_I=0$.
Equations (\ref{mapp0}-\ref{mapp3}) are generalized as follows.

First, let us consider a system having a number of first neighbors per vertex
proportional to $D$.
Let be $\beta_c^{(SG)}$ and $\beta_c^{(F/AF)}$, respectively, 
solutions of the two set of equations (if any)
\begin{eqnarray}
\label{mapp0g}
G_I\left(\{\media{\tanh^2(\beta_c^{(SG)} J_b)}\}\right)&=& 0, 
\qquad D>2,\\
\label{mapp01g}
G_I\left(\{\media{\tanh(\beta_c^{(F/AF)} J_b)}\}\right)&=& 0,
\qquad D>1.
\end{eqnarray} 
Asymptotically, at sufficiently high dimensions $D$,
the critical inverse temperature of the spin glass model $\beta_c$ is given by
\begin{eqnarray}
\label{mappg}
\beta_c=\mathrm{min}\{\beta_c^{(SG)},\beta_c^{(F/AF)}\};
\end{eqnarray} 
and in the paramagnetic phase the following mapping holds 
\begin{eqnarray}
\label{mapp1g}
\left|\frac{\varphi-\varphi_{eff}}{\varphi}\right|=
O\left(\frac{g(\beta^{-1}-\beta_c^{-1})}{D}\right),
\quad \forall \beta\leq \beta_c, 
\end{eqnarray} 
\begin{eqnarray}
\label{mapp2g}
\varphi_{eff}=\frac{1}{l}
\varphi_{I}\left(\{\media{\tanh^{l}(\beta J_b)}\}\right), \qquad D>2^{l-1},
\end{eqnarray}
where
\begin{eqnarray}\fl
\label{mapp3g}
l=\left\{
\begin{array}{l}
2, \quad \mathrm{if}\quad
|\varphi_{I}\left(\{\media{\tanh^{2}(\beta J_b)}\}\right)| \geq 
2|\varphi_{I}\left(\{\media{\tanh(\beta J_b)}\}\right)|, \\
1, \quad \mathrm{if}\quad
|\varphi_{I}\left(\{\media{\tanh^{2}(\beta J_b)}\}\right)| < 
2|\varphi_{I}\left(\{\media{\tanh(\beta J_b)}\}\right)|.
\end{array}
\right.
\end{eqnarray} 
If $\Gamma$ is infinite dimensional only 
in the large sense,
in the above equations $D$ can be settled as infinite and, 
in the limit $\beta\to\beta_c^{-}$,
Eq. (\ref{mapp1g}) holds exactly.

\subsection{Phase diagram: upper critical surface}
If $D\rightarrow \infty$ in the strict sense, Eqs. (\ref{mapp}-\ref{mapp3}) 
and their generalizations, Eqs. (\ref{mappg}-\ref{mapp3g}),
give the exact free energy in the paramagnetic phase ($P$); the
exact critical paramagnetic-spin glass ($P-SG$), $\beta_c^{(SG)}$, and
paramagnetic-$F/AF$ ($P-F/AF$), 
$\beta_c^{(F/AF)}$, surfaces, the
stability of which depends on which of the two ones is the minimum. 
In the case of a homogeneous measure, the suffix $F$ and $AF$ stay
for ferromagnetic and antiferromagnetic, respectively.
In the general case, such a distinction is possible only
in the positive and negative sectors of $\Sigma_I$, whereas, for
the other sectors, we use the symbol
$F/AF$ only to stress that the transition is not $P-SG$.
Finally, the constrains $D>1$ or $D>2$ along the equations  
stress that the mapping is not otherwise defined.
If instead $D\rightarrow \infty$ only in the large sense, 
Eqs. (\ref{mapp}-\ref{mapp3}) 
and their generalizations, Eqs. (\ref{mappg}-\ref{mapp3g}),
are in general exact only in the limit $\beta\to\beta_c^{-}$
and give the exact critical paramagnetic-spin glass 
($P-SG$), $\beta_c^{(SG)}$, and paramagnetic-$F/AF$ ($P-F/AF$), 
$\beta_c^{(F/AF)}$, surfaces.

Notice that at zero field, due to the inequality 
\begin{eqnarray}
\varphi\leq \varphi_I\left(\{\media{\tanh(\beta J_b)}\}\right),
\end{eqnarray}
easily derived from the convexity of the logarithm function and, 
due to the mapping (\ref{mapp1g}),
at least for centered measures,
our estimations for the free energy become trivially zero if
$D\to \infty$ in the strict sense:
\begin{eqnarray}
\label{phi=0}
\lim_{D\to \infty}\varphi=0, \quad \mathrm{for}\quad \beta\leq\beta_c,
\end{eqnarray}
and the basic role of Eqs. (\ref{mapp1}-\ref{mapp3}), 
or (\ref{mapp1g}-\ref{mapp3g}),
is that to show how, in this limit, $\varphi$ 
approaches zero and which are its singularities.

We stress that the presented result is obtained in a replica 
approach method but does not rely on any ansatz
concerning the choice of finding stationary points. In known examples
we have so far considered, Eqs. (\ref{mapp}-\ref{mapp3})  
give results coinciding with those obtained
in the framework of the standard replica approach equipped with
the replica-symmetric ansatz, generally accepted as exact above
the critical temperature. We recall that, above the critical temperature,
the replica symmetric solution
has been proved to be the maximum of the functional appearing in 
the standard replica approach, only in the
SK model and in the ``p-spin'' models \cite{NOTA0}-\cite{Franz}.
We stress also that the mapping is
exact for any case in which $\Gamma$ is infinite dimensional 
and with arbitrary measures including, for example, generalizations 
of the SK model, Bethe lattice models, 
and models defined on generalized tree-like structures.
Finally, Eqs. (\ref{mappg}-\ref{mapp3g}) are useful for more general
models, like anisotropic models and models defined on bipartite lattices.

Several general properties can be immediately derived from the 
critical conditions, Eqs. (\ref{mapp0}-\ref{mapp2}), 
or their generalization, Eqs. (\ref{mapp0g}-\ref{mapp2g}). 
In particular we find as corollaries: 
\begin{list}{}
\item{ \textit{i)} A critical $\beta_c^{(SG)}$ exists 
and is finite $iff$ the related Ising model has a finite 
critical $\beta_c^{(I)}$
\item{ \textit{ii)} 
If an Ising model on $\Gamma$ with a homogeneous coupling $J>0$ 
has a phase transition at some $\beta_c^{(I)}$, then 
the spin glass on $\Gamma$ with a homogeneous ``plus-minus'' 
measure $d\mu(J_b)/dJ_b = \delta(J-J_b)/2+\delta(J+J_b)/2$, will have
a $SG$ transition for $\beta_c^{SG}$ solution of
\begin{equation}
\tanh^2(\beta_c^{(SG)}J)=\tanh(\beta_c^{(I)}J),
\end{equation}}
and, as a consequence, for the plus-minus measure one has always 
\begin{equation}
\label{SGI}
\beta_c^{(SG)}>\beta_c^{(I)}
\end{equation}}
\item{\textit{iii)} A critical $\beta_c^{(F/AF)}$, exists 
and is finite $iff$ for the related Ising model exists a point 
$\{w_b^{(I)}\}\in \Sigma_I$ such that for $w_b^{(I)}\geq 0$ one has
$1-2\int_{-\infty}^0 d\mu_b(J_b)\geq w_{b}^{(I)}$, 
and for $w_b^{(I)}< 0$ one has
$1-2\int_{-\infty}^0 d\mu_b(J_b)\leq w_{b}^{(I)}$, $\forall b\in\Gamma$} 
\item{\textit{iv)} In the space of the parameters of the probability
distribution $d\mu$ (for simplicity here we consider only the case of a
same disorder for any bond), 
possible lines of coexistence $SG-F$ or
$SG-AF$, must intersect the \textit{multicritical} 
points satisfying the systems of equations
given respectively by 
\begin{eqnarray}
\label{coexF}
\left\{
\begin{array}{c}
\mediau{\tanh^{2}(\beta_c J_b)}=w_{F}^{(I)}, \\
\mediau{\tanh(\beta_c J_b)}=w_{F}^{(I)};
\end{array}
\right.
\end{eqnarray}
\begin{eqnarray}
\label{coexAF}
\left\{
\begin{array}{c}
\mediau{\tanh^{2}(\beta_c J_b)}=w_{F}^{(I)}, \\
\mediau{\tanh(\beta_c J_b)}=w_{AF}^{(I)}.
\end{array}
\right.
\end{eqnarray}}
\end{list}

\subsection{Analytic continuations: 
coexistence surfaces and correlation functions}
The statements until now presented are exact.
In this subsection we include the following by-product.
For $D=\infty$, below the critical temperature, 
Eqs. (\ref{mapp1}-\ref{mapp3}) or (\ref{mapp1g}-\ref{mapp3g}) 
are no more valid.
This limit has the peculiar feature that, 
despite the fact that above the 
critical temperature the free energy is exact, in general, it cannot be 
analytically continued to lower temperatures \cite{Palmer2} 
and this happens due to the fact that when $D=\infty$, 
below the critical temperature 
the related Ising model is ill defined or, in other words,
its thermodynamic limit does not exist; the density energy being infinite
for any non zero value of the mean magnetization.

What we argue, instead, is that the analytic continuation
of some physical quantities 
below $\beta_c^{-1}$, even if it is not
rigorous, provides a certain effective approximation.
This is in particular the case 
for the coexistence equations (\ref{coexF}-\ref{coexAF}) for 
evaluating the spin glass-ferromagnetic, ($SG-F$), and 
the spin glass-antiferromagnetic, ($SG-AF$), boundaries.
As we shall show in the examples
we will consider, these crossovers are, roughly speaking, 
close to the Almeida Thouless lines.
Due the fact that the free energy provided by the mapping is
eaxct in all the paramagnetic phase as $D\to\infty$ in the strict sense,
we expect that the above analytic continuation turns out to be better
in these cases, rather than in the cases in which 
$D\to\infty$ only in the large sense.

Similar comments hold for extending the free energy part 
$\varphi$ to the case of arbitrary external fields $h_i$.
As we will discuss in Appendix A, the natural extension
for $\varphi$ consists simply in adding the furhter set of 
arguments $\{\tanh(\beta h_i)\}$ as 
\begin{eqnarray}
\label{mapp2h}
\varphi_{eff}=\frac{1}{l}
\varphi_{I}\left(\{\mediau{\tanh^{l}(\beta J_b)}\};\{\tanh(\beta h_i)\}\right),
\end{eqnarray}
where $l$ is to be determined with the analogous of Eqs. (\ref{mapp3}).

Now, if in infinite dimensions
this extended mapping holds, due to the arbitrariness
of the external fields $h_i$ in $\varphi$, 
by derivation we see that we can even calculate a given 
connected correlation function $g$ starting from the 
knowledge of the corresponding connected correlation function $g_I$ of the
related Ising model. Hence, for example, in infinite dimension,
for a correlation function of
order two at zero external field, we have
\begin{eqnarray}
\label{CORR2}
g^{(2)}(i_1,i_2)\equiv 
\int d\mathcal{P}\left(\{J_b\}\right) 
\left(\mediaT{\sigma_{i_1} \sigma_{i_2}}
-\mediaT{\sigma_{i_1}}\mediaT{\sigma_{i_2}}\right)= 
g_{eff}^{(2)}(i_1,i_2), 
\end{eqnarray}
where $g_{eff}^{(2)}(i_1,i_2)$ is builded as in Eq. (\ref{mapp3})
trough the correlation function $g_I^{(2)}(i_1,i_2)$
\begin{eqnarray}
\label{CORR2I}
g_I^{(2)}(i_1,i_2)\equiv \mediaT{\sigma_{i_1}\sigma_{i_2}}_I
-\mediaT{\sigma_{i_1}}_I\mediaT{\sigma_{i_2}}_I.
\end{eqnarray}
More in general, a connected correlation function of order $k$ 
of the spin glass model is given by
\begin{eqnarray}
\label{CORR}
g^{(k)}(i_1,\ldots,i_k) &\equiv& 
\int d\mathcal{P}\left(\{J_b\}\right) 
\mediaTc{\sigma_{i_1} \ldots \sigma_{i_k}}
= g_{eff}^{(k)}(i_1,\ldots,i_k), 
\end{eqnarray}
where $\mediaTc{\sigma_{i_1} \ldots \sigma_{i_k}}$ represents 
a connected correlation function of order $k$ (see \textit{e.g.} \cite{SHY}), 
and $g_{eff}^{(k)}(i_1,\ldots,i_k)$ is given through
$g_{I}^{(k)}(i_1,\ldots,i_k)$ with the  
analogous of Eqs. (\ref{mapp2}-\ref{mapp3});
$g_{I}^{(k)}(i_1,\ldots,i_k)$ being the connected correlation function
of the corresponding related Ising model 
\begin{eqnarray}
\label{CORRI}
g_{I}^{(k)}(i_1,\ldots,i_k)\equiv \mediaTc{\sigma_{i_1}\ldots \sigma_{i_k}}_I.
\end{eqnarray} 
In this work we will not discuss the quadratic correlation functions
\begin{eqnarray}
\label{CORRIQ}
\int d\mathcal{P}\left(\{J_b\}\right) 
(\mediaTc{\sigma_{i_1} \ldots \sigma_{i_k}})^{2}.
\end{eqnarray} 

\section{Disordered systems vs $n$ interacting Ising like systems}
As is known, a way to carry out Eq. (\ref{logZ}) 
even with a fixed value of $N$ 
(for shortness we will always omit the dependence on $N$), 
consists in applying the replica method
\begin{eqnarray}\fl
\label{REPLICA}
\int d\mathcal{P}\left(\{J_b\}\right)
\log\left(Z\left(\{J_b\}\right)\right) = \lim_{n \rightarrow 0} 
\frac{\int d\mathcal{P}\left(\{J_b\}\right)Z^n-1}{n}=
\lim_{n \rightarrow 0} \frac{Z^{\left(n\right)}-1}{n},
\end{eqnarray} 
where we have introduced the notation
\begin{eqnarray}
\label{Zn0}
Z^{\left(n\right)}\equiv \int 
d\mathcal{P}\left(\{J_b\}\right)Z\left(\{J_b\}\right)^n.
\end{eqnarray} 
Unlike $Z^n$, $Z^{\left(n\right)}$ represents 
the partition function of 
$n$ interacting Ising like systems, \textit{i.e.}, a system 
with $n$ families of
Ising spins $\sigma^{\left(1\right)},\ldots,\sigma^{\left(n\right)}$ 
interacting each one through the usual two-body Ising like
interaction and, among themselves, 
through a non quadratic additional interaction.
For example for a Gaussian measure 
\begin{eqnarray}
\label{dmugauss}
\frac{d\mu\left(J_b\right)}{dJ_b} = \frac{1}{\sqrt{2\pi \tilde{J}^2}}
\exp\left(-\left(J_b-J_0\right)^2/\tilde{J}^2\right),
\end{eqnarray} 
it is easy to recognize that 
\begin{eqnarray}
\label{Zn}\fl
Z^{\left(n\right)} &=& e^{\frac{\beta^2}{2}\sum_{\alpha}\sum_b\tilde{J}^2}
\times 
\sum_{\{\sigma_{b}^{\left(1\right)},\ldots,\sigma_{b}^{\left(n\right)}\}}e
^{ -\beta\sum_{\alpha}\sum_b j_0 \tilde{\sigma}_{b}^{\left(\alpha\right)} + 
\frac{\beta^2}{2}\sum_{\alpha\neq\beta}
\sum_b \tilde{J}^2\tilde{\sigma}_{b}^{\left(\alpha\right)}
\tilde{\sigma}_{b}^{\left(\beta\right)} }, 
\end{eqnarray} 
where the replica indices $\alpha$ and $\beta$ run over $1,\ldots,n$.
A part from constant factors, the case $n=2$ of Eq. (\ref{Zn}) 
corresponds to the Ashkin-Teller model, 
for which in two dimensions some exact results are known 
\cite{ASHKIN}-\cite{MASTROP&G}. 
In particular, it is immediate to solve this model when one chooses 
$J_0=0$, corresponding to the symmetric Gaussian measure. 
In fact, in this simpler case
in Eq. (\ref{Zn}) we are left with the only Ising variable 
$s_b\equiv\tilde{\sigma}_{b}^{\left(1\right)}
\tilde{\sigma}_{b}^{\left(2\right)}$ 
and we are so reduced to a pure Ising sum as follows
\begin{eqnarray}
\label{ISING}
\sum_{\{\sigma_b^{(1)},\sigma_b^{(2)}\}}f(\cdot)=
\sum_{\{\tilde{\sigma}_b^{(1)},\tilde{\sigma}_b^{(2)}\}}f(\cdot),
\end{eqnarray} 
\begin{eqnarray}
\label{ISING1}
\sum_{\{\tilde{\sigma}_b^{(1)},\tilde{\sigma}_b^{(2)}\}}
f(\{\tilde{\sigma}_b^{(1)}\tilde{\sigma}_b^{(2)} \})=
2^{|\Gamma|}\sum_{\{s_{b}\}}f(\{s_{b}\}),
\end{eqnarray} 
$|\Gamma|$ being the number of bonds.
Note however that for $n>2$ the model $Z^{\left(n\right)}$ is no more soluble
even for a symmetric measure. 
This happens due to the impossibility to redefine suitable Ising variables, 
an aspect related to the concept of frustration. 
On the other hand, even in the general case,
it is always possible to find Ising-like contributions in $Z^{\left(n\right)}$.
As we shall see later, 
if the dimension $D$ is sufficiently high, above the critical temperature
the Ising-like contributions become the leading 
part of $Z^{\left(n\right)}$.

It is interesting to note here another aspect of the case $n=2$. 
It has been well established that
the Ashkin-Teller model presents a non universal behavior when 
the sign in front of the quartic coupling $\tilde{J}^2$
in Eq. (\ref{Zn}) is reversed. 
In that case in fact, the specific heat diverges 
with a power law whose exponent
is a continuous function of the quartic coupling. On the other hand such 
a non universal behavior
cannot appear in the ``disordered'' system $Z^{\left(n\right)}$, 
simply because the measure (\ref{dmugauss}) is definite
only for real values of the disorder parameter $\tilde{J}$.
This observation enforce then the idea that at least some features of a  
spin glass could be effectively described through a
suitable Ising model.

\section{Representation of $Z\left(\{J_b\}\right)$ as sum over closed paths}
Let us consider a generic Ising model 
at zero external field with given couplings $\{J_b\}$ 
defined over some set of links $\Gamma$. Note that, since the couplings are
arbitrary, what we will say will be valid, in particular, for the
related Ising model.  
It is convenient to introduce the symbol
\begin{eqnarray}
K_b\equiv \beta J_b.
\end{eqnarray} 
For the partition function it holds the so called 
``high temperature'' expansion
\begin{eqnarray}
\label{Z2}
Z\left(\{J_b\}\right)= \prod_{b\in\Gamma} 
\cosh\left(K_b\right) \sum_{\{\sigma_i\}}
\prod_{b\in\Gamma} \left(1+\tilde{\sigma}_b\tanh\left(K_b\right)\right). 
\end{eqnarray} 
As is known the terms obtained by expansion of the product 
$\prod_{b\in\Gamma} \left(1+\tilde{\sigma}_b\tanh\left(K_b\right)\right)$,
with $k$ bonds proportional to 
$\tilde{\sigma}_{b_1}\tilde{\sigma}_{b_2}\ldots \tilde{\sigma}_{b_k}$, 
contribute to the sum over the spins only
if the set $\gamma\equiv \{b_1,b_2,\ldots,b_k\}$ constitutes 
a closed multipoligon over $\Gamma$ for open or periodic boundary 
conditions,
and a collection of multipoligons and paths, 
whose end-points belong to the boundary of $\Gamma$,
for closed conditions (when all the spins on the boundary are fixed 
to be +1 or -1) (\textit{e.g.} see \cite{GG}); 
in such cases $\tilde{\sigma}_{b_1}\tilde{\sigma}_{b_2}\ldots 
\tilde{\sigma}_{b_k}\equiv 1$
so that Eq. (\ref{Z2}) becomes
\begin{eqnarray}
\label{Z3}
Z\left(\{J_b\}\right)= 2^{N} \prod_{b\in\Gamma} 
\cosh\left(K_b\right) \sum_{\gamma} 
\prod_{b\in \gamma}\tanh\left(K_b\right), 
\end{eqnarray} 
where the sum runs over all the above mentioned paths $\gamma$. 
Note that in the case $\tanh\left(K_b\right)=0$, 
the sum over the paths gives 1, (\textit{i.e.} the contribution
with zero paths must be included). 
Given the set of couplings $\{J_b\}$, in the thermodynamic limit, 
the series in the r.h.s. of Eq. (\ref{Z3}),
normalized to $N$,  will be convergent for values 
of the parameters $z_b=\tanh(\beta J_b)$ sufficiently small,
\textit{i.e.}, inside a suitable set $\mathscr{D}$
whose boundary corresponds to a critical surface $\Sigma$
of the quenched Ising model. Since in Eq. (\ref{Z3}) we have 
a power series, in $z_b$, with positive coefficients, it turns out
that $\mathscr{D}$, 
as anticipated, is a convex set.

Let us consider the important case 
of a homogeneous model: $K_b\equiv K$.
In this case, Eq. (\ref{Z3}) can be written in the form 
of a power series in the single parameter $\tanh(K)$
\begin{eqnarray}
\label{Z3UNIF}
Z = 2^N\cosh^{|\Gamma|}\left(K\right) \sum_{l=0}^{N} 
C_l \tanh^l\left(K\right),
\end{eqnarray} 
where $C_l$ is the number of paths of length $l$.
Note that, for fixed $l$, $C_l$ is a function of the size $N$ and
$C_l \propto N$, so that in the thermodynamic limit
the useful quantity is $c_l\equiv \lim_{N\rightarrow \infty}C_l/N$.
Whereas $C_l$ represents the total number of paths of length $l$ in a given
lattice of finite size $N$, $c_l$ represents the total number of paths
of length $l$ passing trough a single site in an infinite system;
$c_l$ is a growing function of $l$ and for large $l$ this growth
is known to be exponential, both for lattice systems
and for tree-like structures, 
see respectively Refs. \cite{Fisher} and \cite{Ginestra} and
references therein.
As $c_l$ is positive, for positive values of $K$, the series will be
convergent and absolutely convergent if $\tanh(K)<w_{F}^{(I)}$, where 
\begin{eqnarray}
\label{Z3u}
\frac{1}{w_{F}^{(I)}}=\lim_{l\rightarrow \infty}c_l^{\frac{1}{l}},
\end{eqnarray} 
whereas for negative values of $K$, according to the
Leibniz criterion, the series will be convergent
if $w_{AF}^{(I)}<\tanh(K)$, where $w_{AF}^{(I)}$ is defined as
the largest (in modulo) value of $\tanh(K)$ such that
\begin{eqnarray}
\label{Z3l}
\lim_{l\rightarrow \infty}c_l\tanh^l(K)=0.
\end{eqnarray} 
Hence, for $J>0$, in the thermodynamic limit the series 
$\sum_l c_l\tanh^l(\beta J)$ exists for any $\beta<\beta_c^{(F)}$, 
where $\beta_c^{(F)}$, the inverse of the critical temperature, 
is solution of the equation 
$w_{F}^{(I)}=\tanh(\beta_c^{(F)} J)$, and analogously, for $J<0$, the series
will be convergent for any $\beta<\beta_c^{(AF)}$, where 
$\beta_c^{(AF)}$, the antiferromagnetic critical temperature, 
is solution of the equation $w_{AF}^{(I)}=\tanh(\beta_c^{(AF)} J)$. 
In the thermodynamic limit, given a coupling $J$, 
for any value of $\beta$ lesser than the critical one, 
it is possible to define the typical length of the path 
$\bar{l}$ by looking at the distribution 
$p_l\equiv c_l\tanh(\beta J)^l/(\sum_{l'} c_{l'}\tanh(\beta J)^{l'})$.
Note that, as $\beta\to\beta_c^{-}$, the distribution $p_l$ becomes
infinitely flat.
Similarly, in the most general case, in which
at any bond $b$ we can have a different coupling $J_b$, 
the partition function can be rewritten as
\begin{eqnarray}
\label{Z3g}
Z\left(\{J_b\}\right)= 2^{N} \prod_{b\in\Gamma} 
\cosh\left(K_b\right) \sum_{\{l_b=0\}}^N 
C(\{l_b\})\prod_{b\in \gamma}\left(\tanh\left(K_b\right)\right)^{l_b}, 
\end{eqnarray} 
where $C(\{l_b\})$ is the number of paths having $\{l_b\}$ units along
the bonds $\{b\}$, with $l_b=0,1$. Even in this case, 
from the knowledge of
$C(\{l_b\})$, it is possible to calculate the typical length per site
$\bar{l}_b$ along the bond $b$. 
As in the homogeneous case, also in the general case the critical temperature
of the system is determined by the asymptotic behavior of the rescaled
coefficients $c(\{l_b\})\equiv \lim_{N\rightarrow \infty}C(\{l_b\})/N$.
By re-phrasing in terms of paths, in general we have that:
\begin{center}
\textit{The critical behavior of the system is determined by
the paths of arbitrarily large length}
\end{center}
The above observation will be crucial when $\Gamma$ is a generic graph
infinite dimensional only in the large sense.

\begin{figure}[t]
\centering
\includegraphics[width=0.6\columnwidth,clip]{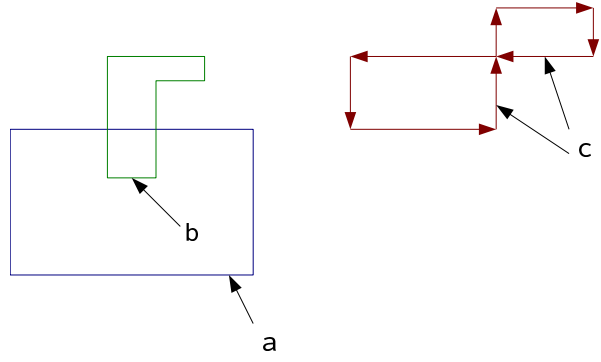}
\caption{Example of one single path $\gamma$ on a 2-dimensional
lattice. In this case $\gamma$ is constituted by three sconnected parts: 
a, b and c.
Note that there is no overlapping between bonds. In the part c of the
path we have drawn arrows to stress that this part of the path is
constituted by only one single connected part, the path c being
connected.}
\label{1_path}
\end{figure}

\section{Averaging over the disorder}
Let us now introduce the universal function $P_{}$  
which takes into account the non trivial part of the high 
temperature expansion
\begin{eqnarray}
\label{P}
P_{}\left(\{z_b\}\right)\equiv \sum_{\gamma} \prod_{b\in \gamma} z_b,
\end{eqnarray} 
and let us average $P_{}$ over the quenched couplings (the disorder)  
\begin{eqnarray}
\label{PA}
P^{(1)}\left(\{F_b^{(1)}\}\right) \equiv \int d\mathcal{P}\left(\{J_b\}\right) 
P_{}\left(\{\tanh(K_b)\}\right), 
\end{eqnarray} 
where we have introduced
\begin{eqnarray}
F_b^{(1)} \equiv \int d\mu_b \tanh(K_b).
\end{eqnarray} 
From the product nature of the distribution 
$d\mathcal{P}\left(\{J_b\}\right)$, 
Eq. (\ref{dP}), it is immediate
to see that $P^{(1)}$ is given in terms of the  
function $P_{}$ through 
\begin{eqnarray}
\label{PA1}
P^{(1)}\left(\{F_b^{(1)}\}\right)=P_{}\left(\{F_b^{(1)}\}\right)=
\sum_{\gamma} \prod_{b\in \gamma} F_b^{(1)}.
\end{eqnarray} 

Later, to evaluate the free energy 
we will need to consider also the averages 
of $P_{I}^n$ for $n\in \mathop{\rm N}$
\begin{eqnarray}
\label{PAn}
P^{(n)}\left(\{F_b^{(1)},\ldots,F_b^{(n)}\}\right) \equiv 
\int d\mathcal{P}\left(\{J_b\}\right) P_{I}^{n}\left(\{\tanh(K_b)\}\right), 
\end{eqnarray} 
where for $m=1,\ldots,n$ we have introduced
\begin{eqnarray}
\label{Fn}
F_b^{(m)} \equiv \int d\mu_b \left(\tanh(K_b)\right)^m.  
\end{eqnarray} 
We note that, according to Eqs. (\ref{JI}-\ref{JIb}),
unlike $P_{}\left(\{\tanh(K_b)\}\right)$,
the function $P_{}\left(\{F_b^{(m)}\}\right)$ is the non trivial
part of the high temperature expansion of the related Ising model
with couplings $\{F_b^{(m)}\}$.

\begin{figure}[t]
\centering
\includegraphics[width=0.6\columnwidth,clip]{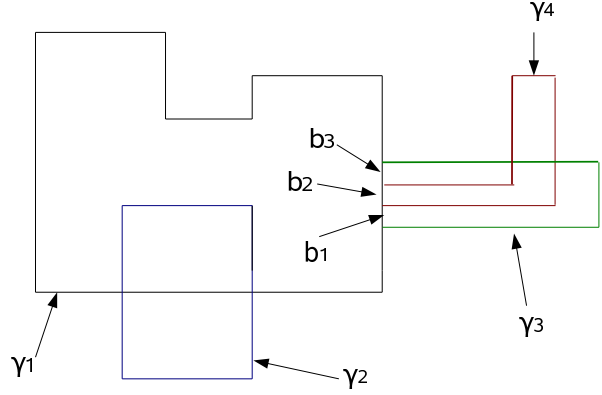}
\caption{A contribution to the summation of Eq. (\ref{PAn1}) with $n=4$.
Here we have: a bond with overlap of order 3, 
$b_2=\gamma_1\cap\gamma_3\cap\gamma_4$;
two bonds with overlap of order 2,
$b_1\cup b_3=\gamma_1\cap\gamma_3$; 
and all the other bonds with no overlap (order 1).
Note that the paths $\gamma_1$ and $\gamma_2$
intersect each other as geometrical objects
but not as sets (for definition, a path $\gamma$ is the union of its bonds).
The same observation holds for the paths 
$\gamma_3\setminus (b_1\cup b_2\cup b_3)$
and $\gamma_4\setminus b_2$.}
\label{4_paths}
\end{figure}

Let us now generalize Eq. (\ref{PA1}) to $P^{\left(n\right)}$. 
From Eqs. (\ref{P}) and (\ref{PAn}) we see that for
$n$ integer we can calculate  $P^{\left(n\right)}$ by summing over $n$ replicas
of paths $\gamma_1,\ldots,\gamma_n$, specifying for any of their
bonds how many overlaps are there with all the other paths
(see Fig. \ref{4_paths}). 
We arrive then at the following expression  
\begin{eqnarray}
\label{PAn1}
P^{\left(n\right)}&=&  
\sum_{\gamma_1,\ldots,\gamma_n} \int d\mathcal{P}\left(\{J_b\}\right) 
\prod_{~~b \in \cap_{l=1}^{n} \gamma_{l}} \tanh^{n}\left(K_b\right) 
\times \nonumber \\
&& \prod_{(i_1)} \prod_{~~b \in \cap_{l=1,l\neq i_1}^{n} \gamma_{l} 
\setminus \gamma_{i_1}} 
\tanh^{\left(n-1\right)}\left(K_b\right) \times \nonumber \\
&& \prod_{\left(i_1,i_2\right)} \prod_{~~b \in \cap_{l=1,l\neq i_1,i_2}^{n} 
\gamma_{l}
\setminus \left(\gamma_{i_1} \cup \gamma_{i_2}\right)} 
\tanh^{\left(n-2\right)}\left(K_b\right) \dots \times \nonumber \\
&& \prod_{(i_n)} \prod_{~~b \in \gamma_{i_n} 
\setminus \left( \cup_{l \neq i_n} \gamma_l \right) } \tanh\left(K_b\right),
\end{eqnarray} 
where, in the product $\prod_{\left(i_1,i_2,\ldots i_k\right)}$, the 
indeces $i_1,i_2,\ldots,i_k$ run 
over the $n!/\left(\left(n-k\right)!k!\right)$ combinations to arrange 
$k$ numbers from the integers $1,\ldots,n$, and with the symbol 
$(i_1,i_2)$ we mean the
couple $i_1,i_2$ with $i_1\neq i_2$ and similarly for 
$\left(i_1,i_2,\ldots i_k\right)$.
From Eq. (\ref{PAn1}) by using Eq. (\ref{dP}) 
and the definitions (\ref{Fn}), we arrive at 
\begin{eqnarray}
\label{PAn2}
P^{\left(n\right)}&=& \sum_{\gamma_1,\ldots,\gamma_n} 
\prod_{~~b \in \cap_{l=1}^{n} \gamma_{l}} F^{\left(n\right)}_{b} 
\times \nonumber \\
&& \prod_{(i_1)} \prod_{~~b \in \cap_{l=1,l\neq i_1}^{n} \gamma_{l} 
\setminus \gamma_{i_1}} F^{\left(n-1\right)}_{b} \times \nonumber \\
&& \prod_{\left(i_1,i_2\right)} \prod_{~~b \in \cap_{l=1,l\neq i_1,i_2}^{n} 
\gamma_{l}
\setminus \left(\gamma_{i_1} \cup \gamma_{i_2}\right)} 
F^{\left(n-2\right)}_{b} \times \dots \times \nonumber \\
&& \prod_{(i_n)} \prod_{~~b \in \gamma_{i_n} 
\setminus \left( \cup_{l \neq i_n} \gamma_l \right) } F^{\left(1\right)}_{b}.
\end{eqnarray} 

\section{Free energy} 
From Eq. (\ref{Z3}) we have
\begin{eqnarray}\fl
\label{logZ1}
\int d\mathcal{P}\left(\{J_b\}\right)\log\left(Z\left(\{J_b\}\right)\right)&=&
\int d\mathcal{P}\left(\{J_b\}\right)
\log\left(2^N\prod_{b\in\Gamma} \cosh(K_b)\right) \nonumber \\ && 
+ \int d\mathcal{P}\left(\{J_b\}\right)
\log\left(\sum_{\gamma} \prod_{b\in \gamma}
\tanh(K_{b})\right),
\end{eqnarray} 
from which, by using Eqs. (\ref{logZ}) and (\ref{dP}) 
in the first term of the r.h.s., we get
\begin{eqnarray}
\label{logZ2}
-\beta F=N\log(2)+
\sum_{b\in\Gamma} \int d\mu_{b} \log\left(\cosh(K_b)\right) + \phi,
\end{eqnarray}
where the non trivial part $\phi$ is given by
\begin{eqnarray}
\label{phi0}
\phi \equiv \int d\mathcal{P}\left(\{J_b\}\right)\log\left(\sum_{\gamma} 
\prod_{b\in \gamma}\tanh(K_{b})\right).
\end{eqnarray} 
With the symbol $\phi_I\left(\{z_b^{(I)}\}\right)$ we will mean the
non trivial part of the free energy of the related Ising model 
\begin{eqnarray}
\label{phiI}
\phi_I \left(\{z_b^{(I)}\}\right)\equiv
\log\left(P_{I}\left(\{z_b^{(I)}\}\right)  \right).
\end{eqnarray} 
The densities of $\phi$ and $\phi_I$ will be indicated as $\varphi$ 
e $\varphi_I$, respectively. 
The free energy term $\phi$ will be obtained in terms of $P^{(n)}$, 
Eq. (\ref{PAn}), via the replica method with the analog of Eq.  
(\ref{REPLICA}):
\begin{eqnarray}
\label{phi1}
\phi = \lim_{n \rightarrow 0} \frac{P^{(n)}-1}{n}.
\end{eqnarray} 

\section{Centered measure}
In this section we will consider the case of a centered measure:
\begin{eqnarray}
\label{dmucen}
\int d\mu_b(J_b)J_b=0,
\end{eqnarray}   
which in particular includes the symmetric case: 
\begin{eqnarray}
\label{dmusim}
d\mu_b(-J_b)=d\mu_b(J_b).
\end{eqnarray}   
For a centered measure we have
\begin{eqnarray}
\label{Fsimm}
F^{\left(2m+1\right)}_{b}= 0.
\end{eqnarray} 
As a consequence, in the high temperature expansion, 
we find the following structure
\begin{eqnarray}
\label{PODD}
P^{(0)}=1,
\end{eqnarray} 
\begin{eqnarray}
\label{PEVEN2}
P^{(2)}=\sum_{\gamma_1=\gamma_2}\prod_{b\in \gamma_1} F^{\left(2\right)}_{b}= 
\sum_{\gamma}\prod_{b\in \gamma} F^{\left(2\right)}_{b}
=P_{}(\{F_b^{(2)}\}), 
\end{eqnarray} 
\begin{eqnarray}
\label{Pcentered}
P^{\left(2n\right)}&=& \sum_{\gamma_1,\ldots,\gamma_{2n}\in \mathscr{E}_{2n}} 
\prod_{~~b \in \cap_{l=1}^{2n} \gamma_{l}} F^{\left(2n\right)}_{b} 
\times \nonumber \\
&& \prod_{\left(i_1,i_2\right)} \prod_{~~b \in \cap_{l=1,l\neq i_1,i_2}^{2n} 
\gamma_{l}\setminus \left(\gamma_{i_1} \cup \gamma_{i_2}\right)} 
F^{\left(2n-2\right)}_{b} \times \dots \times \nonumber \\
&& \prod_{(i_{n-1},i_n)} \prod_{~~b \in \gamma_{i_{n-1}}\cap \gamma_{i_n} 
\setminus \left( \cup_{l \neq i_{n-1},i_n} \gamma_l \right) }
F^{\left(2\right)}_{b},
\end{eqnarray} 
where $\mathscr{E}_{2n}$ is the set that
constrains the sum in the r.h.s. of Eq. (\ref{Pcentered})
to be restricted to combinations of paths
for which any bond $b$ may have only an even number of overlaps
with any other bond.

Unlike the case of Eq. (\ref{PEVEN2}), corresponding to the
centered Ashkin-Teller case, we cannot express the general
$2n$-th term of Eq. (\ref{Pcentered}) in terms of an Ising like
suitable sum as $P_{}(\{F_b^{(m)}\})$. Nevertheless, for any $n$,
in the r.h.s. of Eq. (\ref{Pcentered}) we recognize 
Ising like contributions such as 
\begin{eqnarray}
\label{PEVEN4}
P^{(4)}&=
& \sum_{\gamma}\prod_{b\in \gamma} F^{\left(4\right)}_{b} + 
 3 \sum_{\gamma_1,\gamma_2:\gamma_1 \cap \gamma_2 = \emptyset}
\prod_{b\in \gamma_1} F^{\left(2\right)}_{b} \prod_{b\in \gamma_2} 
F^{\left(2\right)}_{b} + \ldots, 
\end{eqnarray} 
\begin{eqnarray}
\label{PEVEN6}
P^{(6)}&=& \sum_{\gamma}\prod_{b\in \gamma} F^{\left(6\right)}_{b} + 
 15 \sum_{\gamma_1,\gamma_2:\gamma_1 \cap \gamma_2 = \emptyset}
\prod_{b\in \gamma_1} F^{\left(2\right)}_{b} \prod_{b\in \gamma_2} 
F^{\left(4\right)}_{b} \nonumber \\ 
&& + 15 \sum_{\gamma_1,\gamma_2,\gamma_3:\gamma_i \cap \gamma_j = \emptyset, 
i\neq j}
\prod_{b\in \gamma_1} F^{\left(2\right)}_{b} \prod_{b\in \gamma_2} 
F^{\left(2\right)}_{b}
\prod_{b\in \gamma_3} F^{\left(2\right)}_{b}+\ldots,
\end{eqnarray} 
and in general we have
\begin{eqnarray}
\label{PEVEN}
P^{(2n)}&=& \sum_{0 \leq m_1 \leq m_2 \ldots \leq m_n: m_1+\ldots m_n=n} 
C^{(2n)}\left(2m_1,\ldots,2m_n\right) \nonumber \\
&& \times \sum_{\gamma_1,\ldots,\gamma_n: \gamma_i \cap \gamma_j =
\emptyset, i\neq j}~~  
\prod_{b\in \gamma_1} F^{\left(2m_1\right)}_{b} \cdots \prod_{b\in \gamma_n} 
F^{\left(2m_n\right)}_{b}+\ldots,
\end{eqnarray} 
where the coefficient $C^{(2n)}\left(2m_1,\ldots,2m_n\right)$
is given by
\begin{eqnarray}
\label{COMB}
C^{(2n)}\left(2m_1,\ldots,2m_n\right)=
\frac{(2n)!}{\prod_{l=1}^{n}(2m_l)!} \frac{1}{g_{1}! \cdots g_{n'}!},
\end{eqnarray} 
where $n'$ is the number of different values of the $m$'s in the sequence 
$m_1,\ldots,m_n$, and the numbers $g_1,\ldots,g_{n'}$ 
take into account of the degeneracy of the values $m_1,\ldots,m_n$:
\begin{eqnarray}
\label{g}
g_p &=& 
\left\{
\begin{array}{l}
\sum_{l=1}^{n} \delta_{m_p,m_l}, \quad m_p\neq 0 \\
1, \quad m_p=0,
\end{array}
\right.
\end{eqnarray} 
with $p=1,\ldots ,n'$. The number 3 which appears 
in front of the second term of Eq. (\ref{PEVEN4}) counts
the number of possible ways to pair (two to two) 4 paths. 
Similarly in r.h.s. of Eq. (\ref{PEVEN6})
the numbers 15 in front of the second and third terms take into account 
of the possible ways to pair 4 paths and 6 paths, respectively,
from a set of 6 paths. Finally, in Eq. (\ref{COMB}) 
we have written the general form of these combinatorial coefficients. 
However, as we will see, these combinatorial coefficients are of
no importance in the thermodynamic limit.

\begin{figure}[t]
\centering
\includegraphics[width=0.4\columnwidth,clip]{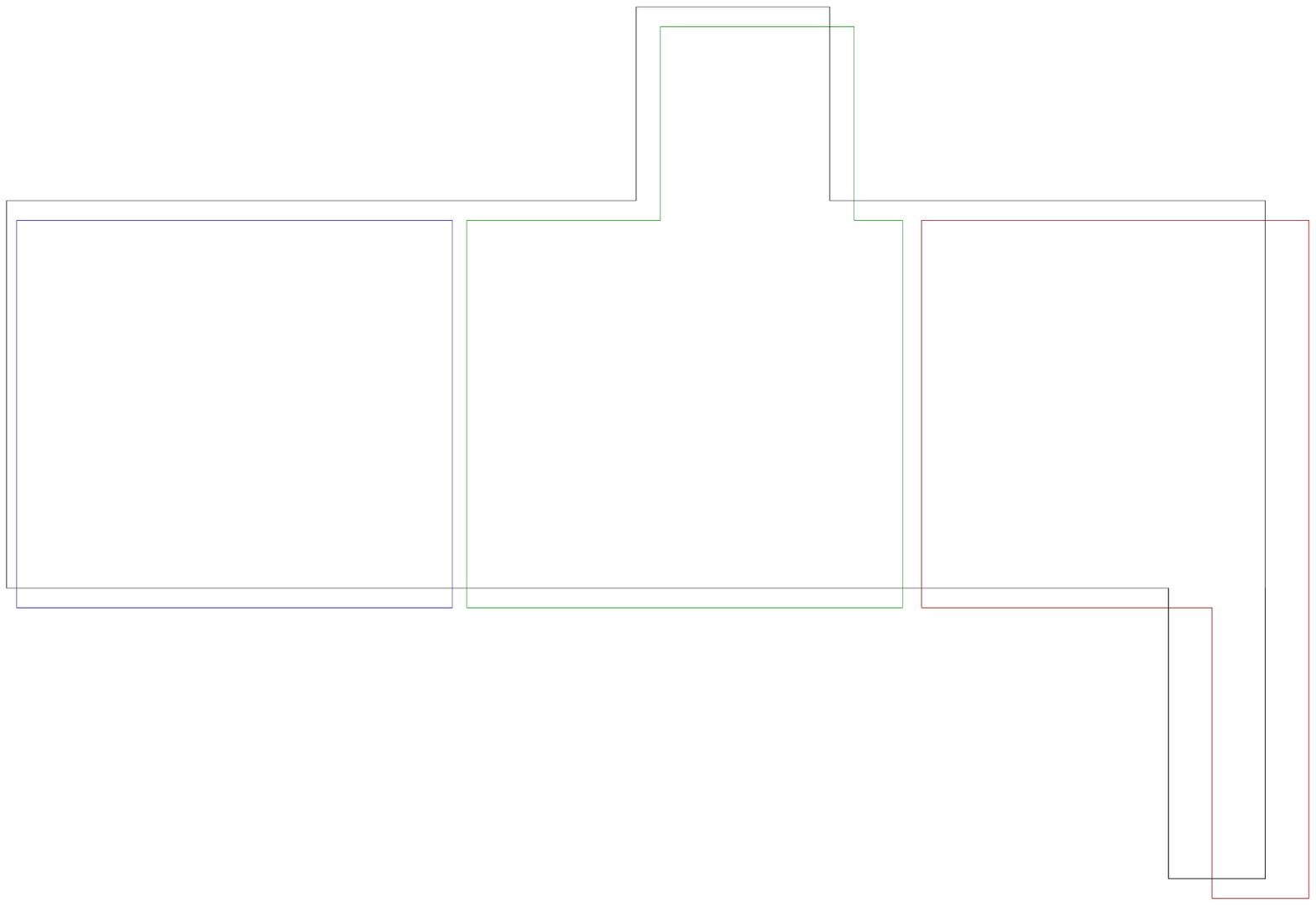}
\caption{ 
A chain of three connected planar paths encapsulated in a larger planar path. 
The four paths 
overlap each other only partially.
The paths in the figure are slightly shifted for visual convenience.}
\label{R}
\end{figure}

In Eqs. (\ref{PEVEN4}-\ref{PEVEN}) we have explicitly written
all the contributions in which any path $\gamma$ coincides with
other paths an even number of times and we have then re-named the
remaining different paths as similarly done 
in the third member of Eq. (\ref{PEVEN2}): $\gamma\equiv \gamma_1=\gamma_2$.
All these terms, up to the constrain that the re-defined paths $\gamma$'s
cannot have common segments, are Ising like terms, whereas
in Eqs. (\ref{PEVEN4}-\ref{PEVEN})
we have left dots, $\ldots$, for indicating the other contributions which
come from more complicated combinations of paths in which two or
more paths $\gamma$ overlap by themself only partially 
(\textit{i.e.} not entirely two to two) as
shown for example in Fig. \ref{R}.
These latter contributions cannot be expressed as Ising like terms.

\subsection{Approximation in high dimensions}
\label{HD}
Up to now we have not yet introduced any approximation in our scheme.
Unfortunately, the equation (\ref{PEVEN}) for $P^{(2n)}$ presents  
elements of intractability which avoid an exact calculation even 
in two dimensions where,
on the contrary, the exact solution is possible for the pure Ising model.
These difficulties are of two kind.
First, in the Ising like terms, 
the re-named $n$ paths $\gamma$ 
cannot have common segments or, in other words, 
we have to sum over the ensemble of $n$ non overlapping random walks.
Second, we do not have any knowledge about the non Ising like terms.

Nevertheless, we can show that, as the dimension $D$ of the system
grows, neglecting either the constrain for the non overlapping
random walks and the non Ising like terms, implies a 
smaller and smaller error in evaluating the general 
expression (\ref{PEVEN}) which, 
in the limit $D\rightarrow \infty$, becomes exact.

In the following we will make use of the fact that
for finite and positive values of $m$ we have 
$O(F_b^{(2m)})\sim O(F_b^{(2)})^{m}$. In particular,
for the $\pm J$ distribution we have $F_b^{(2m)}=(F_b^{(2)})^{m}$. 
We anticipate that, in the limit of infinite dimension, any finite difference
between $F_b^{(2m)}$ and $(F_b^{(2)})^{m}$ becomes irrelevant for
the mapping to be valid.  
For $\Gamma$, to be specific, we will consider a $D$-dimensional
hypercube lattice. However, as will be evident later, similar arguments
can be repeated for any systems of links $\Gamma$ 
infinite dimensional in the large sense (see Sec. 3),
as happens, but not only, 
in the case of generalized tree-like structures.

\begin{figure}[t]
\centering
\includegraphics[width=0.2\columnwidth,clip]{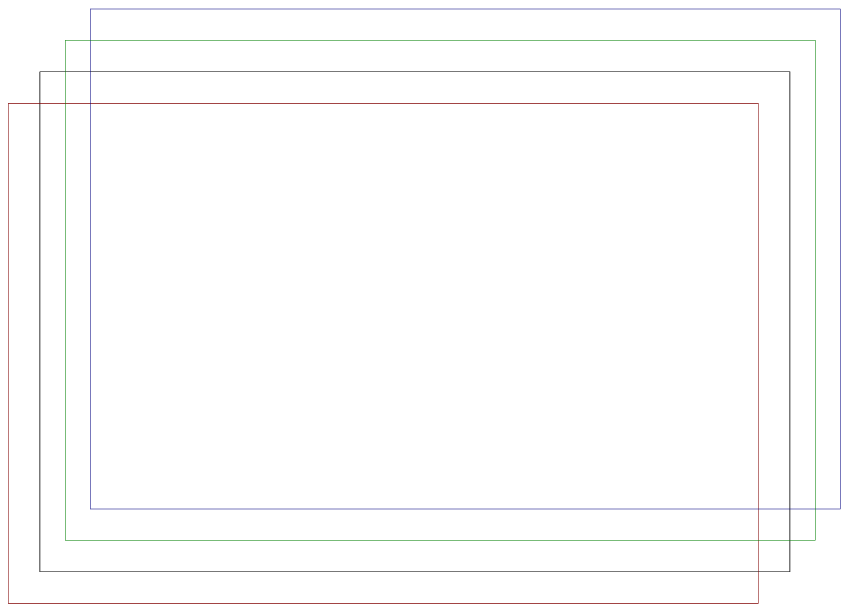}
\caption{ 
Schematic example of 
four completely overlapping planar paths. 
The paths in the figure are slightly shifted for visual convenience.}
\label{4_overlaps}
\end{figure}

\begin{figure}[t]
\centering
\includegraphics[width=0.4\columnwidth,clip]{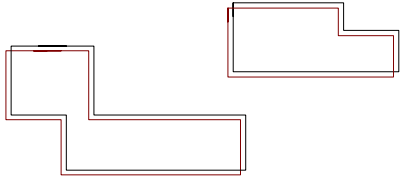}
\caption{ 
Schematic example of 
two couples of two completely overlapping planar paths.
The paths in the figure are slightly shifted for visual convenience.}
\label{2_and_2}
\end{figure}

For the moment, let us consider for simplicity only planar connected paths.
Let us start with $n=4$. We have to sum over all contributions 
that belong to the set $\mathscr{E}_{4}$. From the set of all the possible
planar connected paths, we have to draw out 
four arbitrary paths and combine them in any possible way compatible with
the set $\mathscr{E}_{4}$. We find useful to decompose this set as follows
\begin{eqnarray}
\label{E4}
\mathscr{E}_{4}=\mathscr{E}_{4,4}\cup\mathscr{E}_{2,4}\cup\mathscr{R}_{4},
\end{eqnarray}  
where $\mathscr{E}_{4,4}$ is the set of 4 coinciding paths, \textit{i.e.}
the set of all one replica paths, 
see Fig. \ref{4_overlaps};
$\mathscr{E}_{2,4}$ is the set of all paths
overlapping in couples (two to two), see Fig. \ref{2_and_2};
and $\mathscr{R}_{4}$ is the rest, \textit{i.e.}
the set of all paths in $\mathscr{E}_{4}$
which overlap each other only partially, see Fig. \ref{R}.

We now observe that in general, if $D$ is the dimension, and $D>2$, 
at any point of our lattice system 
we have $D$ orthogonal planes 
where anyone of the 4 planar paths can be arranged.
On the other hand, whereas any given element of
$\mathscr{E}_{4,4}$ or $\mathscr{R}_{4}$ 
lies necessarily on one single plane,  
we can generate an element of $\mathscr{E}_{2,4}$
by arranging two arbitrary paths,
even in two separated planes, for a 
total of $D(D-1)+D$ combinations. 
Therefore, we have 
\begin{eqnarray}
\label{E4e}
\frac{|\mathscr{E}_{4}|-|\mathscr{E}_{2,4}|}
{|\mathscr{E}_{4}|}=O\left(\frac{1}{D-1}\right).
\end{eqnarray}  
Note also that, up to the constrain $\gamma_1\cap\gamma_2=\emptyset$,
the set $\mathscr{E}_{2,4}$
corresponds to the range of the summation 
in the second term of Eq. (\ref{PEVEN4}).

Analogously, for $n=6$, we have to consider the following decomposition
\begin{eqnarray}
\label{E6}
\mathscr{E}_{6}=\mathscr{E}_{6,6}\cup\mathscr{E}_{4,6}\cup\mathscr{E}_{2,6}
\cup\mathscr{R}_{6},
\end{eqnarray}  
and we see again that the elements of the set $\mathscr{E}_{2,6}$
can be arranged in $D(D-1)(D-2)+O(D(D-1))$ 
ways, whereas the elements of the other sets can be arranged at most
in $D(D-1)$ combinations. Therefore we have
\begin{eqnarray}
\label{E6e}
\frac{|\mathscr{E}_{6}|-|\mathscr{E}_{2,6}|}
{|\mathscr{E}_{6}|}=O\left(\frac{1}{D-2}\right).
\end{eqnarray}  
Note that, up to the constrain for the non overlapping of the re-named
paths, the set $\mathscr{E}_{2,6}$ 
corresponds to the range of the 
summation in the third term of Eq. (\ref{PEVEN6}).

We can repeat the above argument for any positive integer $n$.
We always arrive at the conclusion that
the Ising like contributions obtained by pairing  
in all the possible ways two paths are the leading contributions 
of the set $\mathscr{E}_{n}$ as follows
\begin{eqnarray}
\label{Ene}
\frac{|\mathscr{E}_{2n}|-|\mathscr{E}_{2,2n}|}
{|\mathscr{E}_{n}|}=O\left(\frac{1}{D-n+1}\right).
\end{eqnarray}  
Up to the constrain
that the re-named paths must not overlap among themselves, 
the set $\mathscr{E}_{2,2n}$ corresponds to the range of
the summation over paths in Eq. (\ref{PEVEN}).
On the other hand, by applying the same argument of the dimensionality,
we see that neglecting the constrain implies a vanishing error for large $D$,
being
\begin{eqnarray}
\label{Ene1}
\frac{|\mathscr{E}_{2,2n}|-
\sum_{\gamma_1,\ldots,\gamma_n: \gamma_i \cap \gamma_j =
\emptyset, i\neq j} 1}
{|\mathscr{E}_{2,2n}|}=O\left(\frac{1}{D-n+1}\right).
\end{eqnarray}  

Finally, it is not difficult to convince oneself that the a very similar 
argument can be applied 
also for general paths, non connected and non planar. In conclusion,
taking into account Eqs. (\ref{Ene}) and (\ref{Ene1}), the general
expression (\ref{PEVEN}), and the fact that
$O(F^{(2m)})\sim O(F^{(2)})^{m}$, we arrive at
\begin{eqnarray}
\label{PEVEN03V}
P^{(2n)} &=& \frac{(2n)!}{2^n n!} P_{}^{n}\left(\{F_b^{(2)}\}\right)
+ O\left(\frac{P^{(2n)}}{D-n+1}\right).
\end{eqnarray}  

Furthermore, since for $\beta^{-1}>>(\beta_c^{(I)})^{-1}$ 
the typical length $\bar{l}(\beta)$ goes to zero, we can improve
the description of the above error including an unknown
bounded function $g(x)$ of order 1 which goes to zero for
$\beta^{-1}>>(\beta_c^{(I)})^{-1}$ and takes into account the 
fact that in such a limit Eq. (\ref{PEVEN03V}) becomes
a trivial identity, being both $P^{(2n)}$ and 
$P_{}^{n}\left(\{F_b^{(2)}\}\right)$ equal to 1 (as we shall see shortly
the coefficient in front of $P_{}^{n}$ does not play any role in the
thermodynamic limit).
Therefore we can replace Eq. (\ref{PEVEN03V}) with the finer one
\begin{eqnarray}
\label{PEVEN03}
P^{(2n)} &=& \frac{(2n)!}{2^n n!} P_{}^{n}\left(\{F_b^{(2)}\}\right)
+ O\left(\frac{P^{(2n)}g(\beta^{-1}-(\beta_c^{(I)})^{-1})}
{D-n+1}\right).
\end{eqnarray}  
By exploiting this approximation, we are now able
to calculate the free energy term $\phi$ via the replica 
formula of Eq. (\ref{phi1}) or, more precisely, via the following limit
\begin{eqnarray}
\label{phi20}
\phi = \lim_{n \rightarrow 0} \frac{P^{(2n)}-1}{2n}.
\end{eqnarray} 

By using the relation
\begin{eqnarray*}
(2n)! &=& 2^n n! (2n-1)!!,
\end{eqnarray*}  
and the Gamma function property
\begin{eqnarray*}
(2n-1)!! = \frac{\Gamma\left(n+\frac{1}{2}\right)}
{\Gamma\left(\frac{1}{2}\right)} 2^n,
\end{eqnarray*}  
form Eq. (\ref{PEVEN03}) we arrive at
\begin{eqnarray}
\label{PEVEN04}
P^{(2n)} &=& \frac{\Gamma\left(n+\frac{1}{2}\right)}
{\Gamma\left(\frac{1}{2}\right)} 2^n
P_{}^{n}\left(\{F_b^{(2)}\}\right)
+ O\left(\frac{P^{(2n)}}{D-n}\right).
\end{eqnarray}  
Finally, by taking the derivative with respect to $n$ we get 
\begin{eqnarray}
\label{phi2}
\phi &=& \lim_{n \rightarrow 0} \frac{P^{(2n)}-1}{2n}=
\frac{1}{2}\log\left(P_{}\left(F_b^{(2)}\right)\right) 
+ O\left(\frac{\phi}{D}\right) \nonumber \\
&& + \frac{1}{2}\log(2)+ 
\frac{1}{2} \int_{0}^{\infty}dt \frac{\log\left( t \right)}
{t^{\frac{1}{2}} e^t},
\end{eqnarray} 
where we have used
\begin{eqnarray}
\label{THIRDTERM}
\lim_{n \rightarrow 0} \frac{\partial}{\partial_n}\log
\left(\Gamma\left(n+\frac{1}{2}\right)\right)=
\int_{0}^{\infty}dt \frac{\log\left( t \right)}{t^{\frac{1}{2}} e^t}.
\end{eqnarray} 
Equation (\ref{phi2}) can also be written as
\begin{eqnarray}
\label{phi3}
\phi &=& 
\frac{1}{2}\phi_I\left(\{F_b^{(2)}\}\right) 
+ O\left(\frac{\phi}{D}\right)
+ \frac{1}{2}\log(2)+ 
\frac{1}{2} \int_{0}^{\infty}dt \frac{\log\left( t \right)}
{t^{\frac{1}{2}} e^t}.
\end{eqnarray} 
Finally, we note that in the thermodynamic limit the last two terms 
in the above equation are of no importance, 
so that, for $N$ large and by using Eq. (\ref{logZ2}), 
for the free energy $F$ we find the following expression  
\begin{eqnarray}
\label{logZ3}
-\beta F&=&\sum_{b\in\Gamma} \int d\mu_{b} \log\left(2\cosh(K_b)\right) + 
\frac{1}{2}\phi_I\left(\{F_b^{(2)}\}\right) \nonumber \\
&& + O\left(\frac{\phi}{D}\right), \quad D>2,
\end{eqnarray}
\begin{eqnarray} 
\label{Fb2}
\quad F_b^{(2)}&=&\int d\mu_{b} \tanh^2(K_b),\quad b\in\Gamma.
\end{eqnarray}
Equation (\ref{logZ3}-\ref{Fb2}) has been derived starting from
the high temperature expansion of the related Ising model, therefore,
taking into account that $\mathscr{D}_I$ is convex,
it holds for any temperature $\beta^{-1}$ 
such that $\media{\tanh^2(\beta J_b)}\leq w_{b}$. 
Equation (\ref{logZ3}-\ref{Fb2}) tells us that for such temperatures, 
the free energy of an Ising spin glass model defined 
over a $D$-dimensional set of links $\Gamma$, with $D$ large and 
in the presence of a centered disorder, can be effectively
expressed in terms of the free energy of an Ising model defined over
the same set $\Gamma$ in which
the parameters of the high temperature expansion 
are replaced by the following effective substitution
\begin{eqnarray}
\label{subst}
z_b=\tanh(K_b) \rightarrow F_b^{(2)}=\int d\mu_b \tanh^2(K_b),\quad b\in\Gamma.
\end{eqnarray} 
In particular, if the set of equations $\media{\tanh^2(\beta J_b)}= w_{b}$
admit a solution for some $\beta_c$, $\beta_c^{-1}$ will be
the critical temperature of a $P-SG$ transition.


\section{Generalization to non centered measures}
If $F_b^{(2m+1)}$ is no more 0, in constructing $P^{(2n)}$,
besides the terms in the set $\mathscr{E}_{2n}$, in which appear
only terms involving $F_b^{(2m)}$, $m=1,\ldots,n$, we have to consider
all the other terms belonging to the complement of $\mathscr{E}_{2n}$,
$\bar{\mathscr{E}}_{2n}$, \textit{i.e.}, 
all the families of contributions in which among the $2n$
paths, there is at least one path in which one or more bonds do not
overlap any other bond of the other paths or, if there is an overlap,
it is an odd overlap, \textit{i.e.}, an odd number of paths
overlap over the same bond. 
Even in this case we can single out the Ising like terms either
with even or odd overlaps of the bonds. 
To this aim we find useful to decompose the whole set of $2n$ paths as:
\begin{eqnarray}
\label{paths}
{\mathscr{E}}_{2n}\cup\bar{\mathscr{E}}_{2n}=\mathscr{F}_{0}\cup\ldots\cup
\mathscr{F}_{k}\cup\ldots\cup\mathscr{F}_{n}\cup\mathscr{R}_n
\end{eqnarray} 
where, for each $k$, $\mathscr{F}_{k}$ is the set in which any element is
constituted by $2n$ paths in which $2k$
of them overlap each other only an even number of times and
the remaining $2n-2k$ paths overlap each other only an odd number of times,
whereas $\mathscr{R}_n$ is the set of all the possible non Ising like terms.

\subsection{Approximation in high dimensions}
When the dimension $D$ is sufficiently high, 
in each subset $\mathscr{F}_{k}$ we can
repeat the same argument as in the previous section as follows.
Among the $2k$ paths (only even overlaps), the most important
terms are those with minimum non zero overlap, \textit{i.e.},
those terms which give a factor $\propto P_{}^{k}(\{F_b^{(2)}\})$, and,
similarly, among the remaining $2n-2k$ paths (only odd overlaps),
the most important terms are those with no overlap, \textit{i.e.},
those terms which give a factor $\propto P_{}^{2n-2k}(\{F_b^{(1)}\})$.
Note that the proof for the second  statement
runs as in subsection (\ref{HD}), the only difference now is that 
it works for paths having odd overlaps, so that, in this case,
the minimum number of overlapping paths is 1, as opposed to 2 in 
subsection (\ref{HD}), and, as a consequence, 
the relative error in neglecting the other
terms goes as $1/(D-2n+1)$, instead of $1/(D-n+1)$.
Furthermore, the minimal dimension to apply the argument for paths
with only odd overlaps is not $D>2$, but $D>1$.

For what concerns the set $\mathscr{R}_n$, we can neglect these
non Ising like contributions also by applying almost the same
argument of the subsection (\ref{HD}). The only difference now is that
these non Ising like terms involve also paths having both
even and odd overlaps. 

In conclusion, we arrive at the following expression for $P^{(2n)}$ 
\begin{eqnarray}\fl
\label{P2ng}
P^{(2n)}=\sum_{k=0}^n P_{}^{k}(\{F_b^{(2)}\})
P_{}^{2n-2k}(\{F_b^{(1)}\})\frac{(2k)!}{2^k k!}
\left(
\begin{array}{c}
2n \\
2k
\end{array}
\right)+O\left(\frac{P^{(2n)}}{D}\right), 
\end{eqnarray} 
where the factor ${(2k)!}/{(2^k k!)}$, as before,
takes into account the number of ways to pair $2k$ paths, whereas
the binomial coefficient takes into account the number of ways to
choose $2k$ paths from a total of $2n$.
Yet, as in the previous section, these combinatorial coefficients
are of no importance in the thermodynamic limit,
where, both the factors, $P_{}(\{F_b^{(2)}\})$ and
$P_{}(\{F_b^{(1)}\})$, grow exponentially in $N$, so that, in this 
limit, only the two leading terms, $P_{}^n(\{F_b^{(2)}\})$ 
and $P_{}^{2n}(\{F_b^{(1)}\})$, are important and, by taking the logarithm
of this sum in Eq. (\ref{phi20}), the final formulas given in Sec. 4 follow.

\section{General graphs}
\label{GRAPH}
In the previous sections we have proved the mapping 
by providing a simple argument which applies when the dimension
$D$ is sufficiently high. Roughly speaking, the key point is that, 
if $D$ is the number of axis passing through a site, by choosing at random 
two of them (or replica), the probability that they coincide (or overlap), 
goes to zero as $1/D$. 
For pedagogical reasons we have used 
this argument by referring to models whose set of links $\Gamma$ is
defined over a $D$ dimensional hypercube lattice $\Lambda$,
where $D$ is related to the number of first neighbors, $2D$.
In these models it is easier to visualize the axis.
However, as we will see soon, little changes are involved in 
the proof if we consider a set $\Gamma$ infinite dimensional 
in the large sense.
Even for these systems,
we can always find the analogous of the number $D$ of independent
axis per site and look at the situation when $D \to \infty$. 
%

Let us see now, more specifically, an Ising model defined over a Cayley tree of
coordination number $q=k+1$. It is constructed as follows: one starts from
a vertex root `$0$' and adds $q$ points all connected to $0$. This set of $q$
points represents the first shell of $q$ sites. The second shell is 
instead obtained connecting each one of these sites to new $k=q-1$ 
points, and so on for the successive shells ($k$ is then the branching factor).
  
This lattice is a tree and, therefore,
if we do not close in some way the boundaries,
or if we do not broke in some minimal way
the symmetry up-down, as happens on a hypercube lattice,
the non trivial part of the free energy $\varphi_I$ is zero,
and no phase transition is possible.
Note that, since a Cayley tree depends heavily and non trivially 
on the boundary conditions, what is usually studied 
is a Bethe lattice, consisting in a subset of the Cayley tree
infinitely far from the boundary. 
Yet, even with such a definition, it is known that,
unlike the Ising model over the Bethe lattice,
the spin glass model over the Bethe lattice remains
still dependent on the kind of the boundary coinditions
imposed on the sites of the outer part of the Cayley tree. 
It is known in particular that, even though in this model there is 
a singularity in the free energy, such a singularity
implies a ``true spin glass phase transition'' 
(\textit{i.e.}, with the same qualitative picture 
as the replica symmetry breaking in the SK model \cite{Parisi})
only under certain boundary conditions
\cite{Lai}. 
However, such a distinction for us is not important;
our aim here is limited to find only the singularities of the model,
regardless on the fact that such singularities may or may not
imply a true spin glass phase transition.

Let us consider a centered measure and start from the general 
exact representation (\ref{Pcentered}).
As we will see, unlike hypercube lattices, 
if we analyze the Ising-like contributions 
for a Bethe lattice, we do not reach exactly the 
Eqs. (\ref{PEVEN4}-\ref{PEVEN}), but a slight modified version of them,
which, however, in the thermodynamic limit, coincide.
Let us look at the set of all the possible paths starting 
from the root 0 and arriving at some
infinitely far boundary. Note that, in a Bethe lattice, regular or not,
such a set coincides with the set of all possible paths.
Furhtermore, as will be clear soon, in a Bethe lattice, and more
in general in a tree, due to the absence of loops,
the effectes of non-Ising like terms are irrelevant,
being limited to an overlapping of a finite number of bonds.
 
\begin{figure}[t]
\centering
\includegraphics[width=0.4\columnwidth,clip]{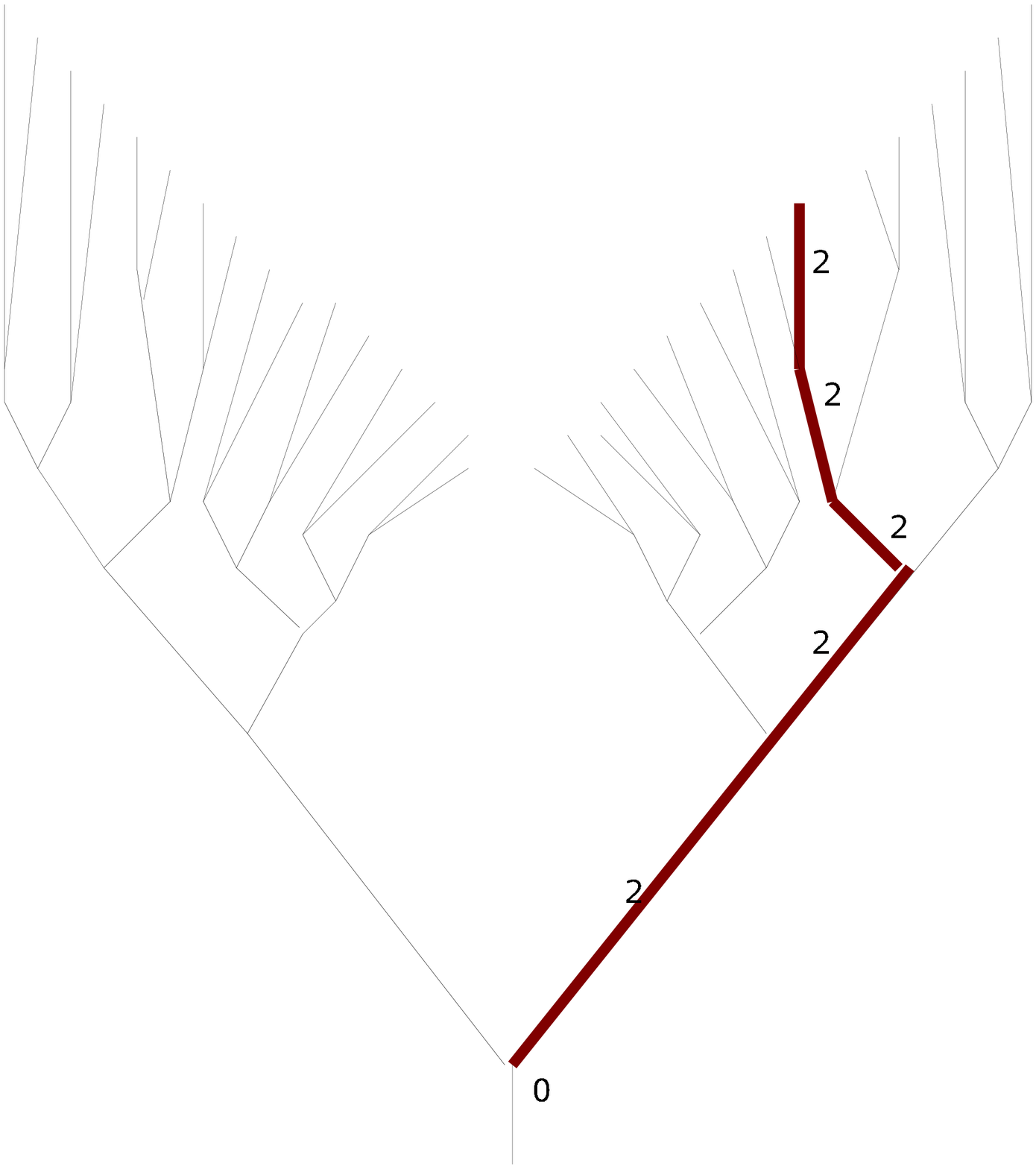}
\caption{ 
An example of a Bethe lattice with coordination number $q=3$ ($k=2$).
The path of greater thickness represents two completely
overlapping trajectories $\gamma=\gamma_1=\gamma_2$ going toward 
the boundary of the lattice.}
\label{simple_tree_2}
\end{figure}

\begin{figure}[t]
\centering
\includegraphics[width=0.4\columnwidth,clip]{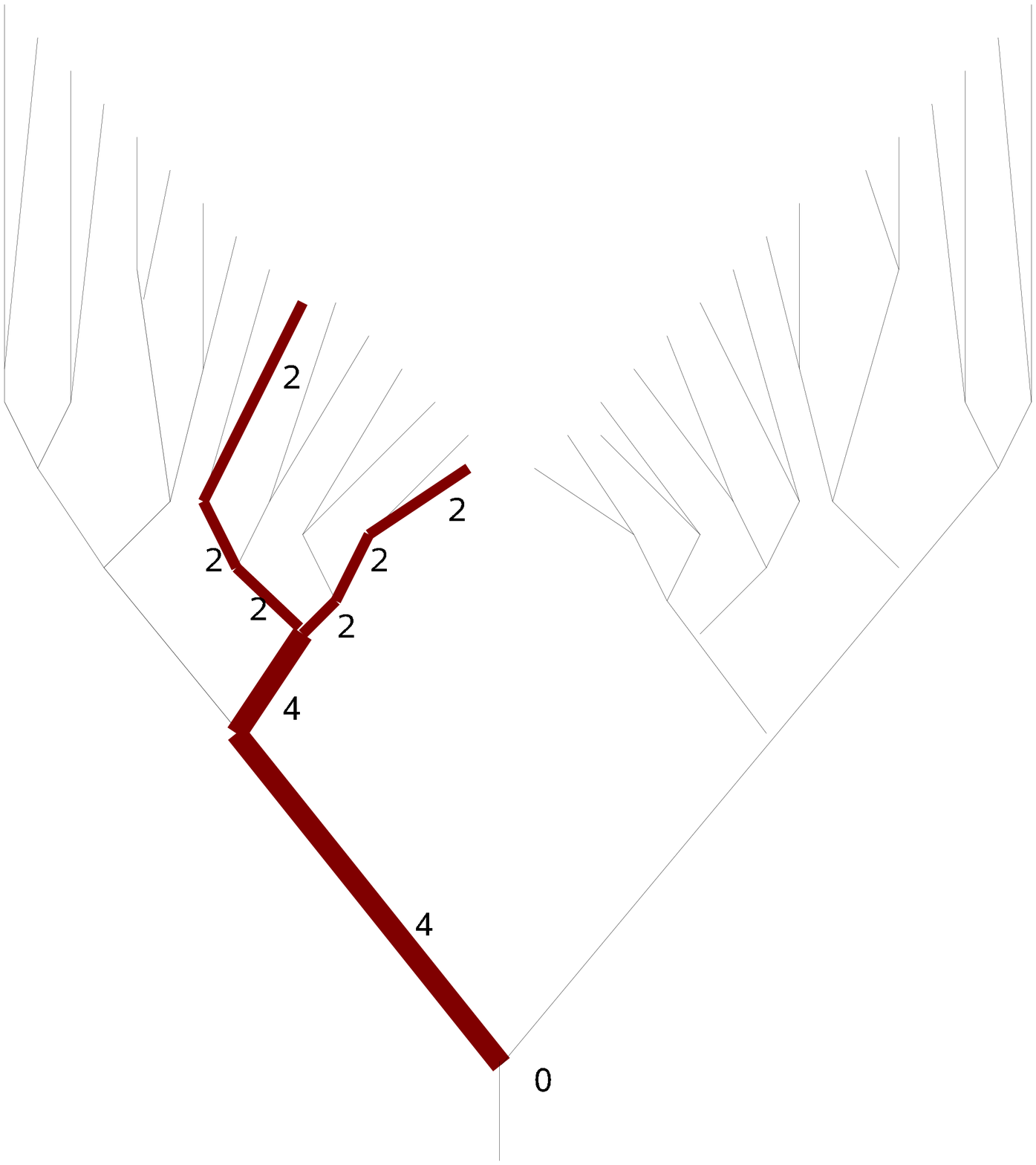}
\caption{ 
Bethe lattice with $q=3$ ($k=2$) 
with four trajectories overlapping four times along the bonds with label
``4'' and overlapping 2 to 2 along the bonds with label ``2''.} 
\label{simple_tree_4}
\end{figure}

It is immediate to see that the term $P^{(2)}$ remains formally 
as Eq. (\ref{PEVEN2}) (see Fig. \ref{simple_tree_2}).
Let us calculate $P^{(4)}$. Due to the absence of loops, 
if 4 replica paths starting from 0 coincide for $l$
bonds, after a splitting at the bond $l+1$ in 2 branchs of 2 
coinciding paths, the branchs will not overlap each other anymore 
(see Fig. \ref{simple_tree_4}).
Therefore, taking into account that the total number 
of paths of length $l$ starting from 0, is $q k^{l-1}$,
the equivalent of Eq. (\ref{PEVEN4}) for a Bethe lattice with $n$
shells becomes 
\begin{eqnarray}
\label{PEVEN4T}
P^{(4)}&=&
q\sum_{l=0}^{n} k^l \prod_{r=1}^{l} F^{(4)}_{b_r}  
3\sum_{\gamma_1,\gamma_2}
\prod_{b\in \gamma_1\setminus \cup_{r=1}^l b_r} F^{(2)}_{b} 
\prod_{b\in \gamma_2\setminus \cup_{r=1}^l b_r} F^{(2)}_{b}, 
\end{eqnarray} 
where $\{b_r\}_{r=1}^{n}$, is a sequence of $n$ successive bonds
starting from 0. 
On the other hand, in the thermodynamic limit, $n\to \infty$, the constrains 
$\setminus \cup_{r=1}^l b_r$ in Eq. (\ref{PEVEN4T}) become negligible and
we get
\begin{eqnarray}
\label{PEVEN4T2}
P^{(4)}&=&
3 q\sum_{l=0}^{\infty} k^l 
\prod_{r=1}^{l} F^{\left(4\right)}_{b_r} 
P_{}^{2}(\{F_b^{(2)}\}).
\end{eqnarray} 
Similar expressions can be derived for any $n$.
Even if these expressions 
become quite involved for increasing values of $n$,
in the thermodynamic limit we are always left with a general
form of the type
\begin{eqnarray}
\label{PEVENT2}
P^{(2n)}&=& B_n P_{}^{n}(\{F_b^{(2)}\}),
\end{eqnarray} 
where $B_n$ is a suitable constant which takes into account
the combinatorics, Eq. (\ref{COMB}), and integers
powers of the sums $q\sum_{l=0}^{\infty} k^l 
\prod_{r=1}^{l} F^{\left(m\right)}_{b_r}$, for $4\leq m \leq n$.
Since both of these factors do not grow with the size
of the system, as for the previous case of subsection 9.1,
we arrive at the same conclusion of Eq. (\ref{logZ3}) 
(with $D=\infty$ in this case).

\begin{figure}[t]
\centering
\includegraphics[width=0.4\columnwidth,clip]{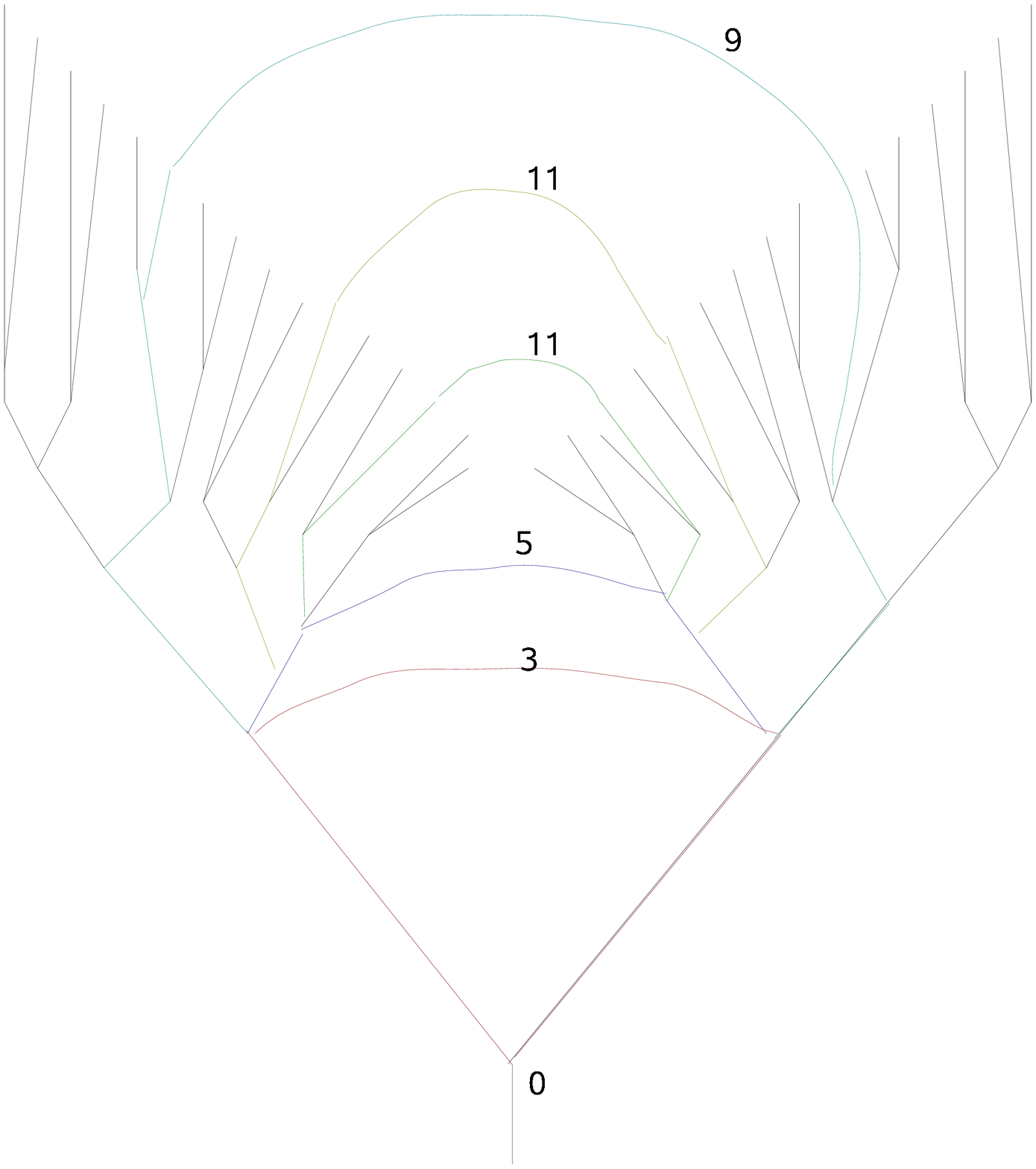}
\caption{ 
A generalized tree-like structure with loops obtained starting from a 
a Bethe lattice having previously $q=3$ ($k=2$).
The drawn loops pass through the root vertex 0.
The numbers over the loops are the corresponding lengths.} 
\label{simple_tree_loop}
\end{figure}

A similar argument can be repeated for tree-like structures 
and for graphs in which there is at most a finite number of 
loops per vertex. 
Unlike the pure Cayley tree now, we have to consider also a finite 
number of closed paths (see Fig. \ref{simple_tree_loop}),
but the construction of the terms $P^{(2n)}$ runs in the same way.
Finally, we see that, as anticipated in Sec. 3, 
we can consider even more complex non tree-like structures
in which the number of loops per vertex is not finite;
the only condition we need being that the paths, two at two, share  at most
a finite number of bonds.
The only problem with such complex structures
is that the related Ising models are hardly solvable. 
An exception to this is provided by the fully connected graph, 
\textit{i.e.}, the SK model.

\section{Applications to finite dimensional models}
We now apply the mapping 
to some finite dimensional cases and compare them with known results. 
Hereafter we will consider mainly homogeneous models, $d\mu_b \equiv d\mu$, 
so that, in particular, $F_b^{(2)}\equiv F^{(2)}$. 
Furthermore in this section we will consider only centered measure
and hypercube lattices.
For the Ising like term, $\varphi_I(F^{(2)})$, in one and two dimensions,
we can use its analytical knowledge coming from the exact solution of the Ising
model, whereas, in higher dimensions, we can use for it the knowledge
coming from numerical methods.  
Clearly, since our approach takes into account only the leading term
of a $1/D$ expansion, in low dimensional systems, we expect poor results, but
it is however interesting to see even in these cases how
our approach works. We will force the study even 
in the cases $D=1$ and $D=2$ which can be seen as singular points
where the mapping is not definite. 

\subsection{One dimensional case}
For $D=1$, if one neglects irrelevant boundary 
effects, there are no closed paths so that from a direct
calculation one sees that $\varphi$ is exactly 0. 
The one dimensional case is therefore trivial. 
Nevertheless, it is worth to observe that, for the same reason, even
$\varphi_I$ is exactly zero, so that for $D=1$, the mapping,
even if not definite, turns out to be exact.

\subsection{Two dimensional case}\label{Two}
\begin{figure}[t]
\centering
\includegraphics[width=0.7\columnwidth,clip]{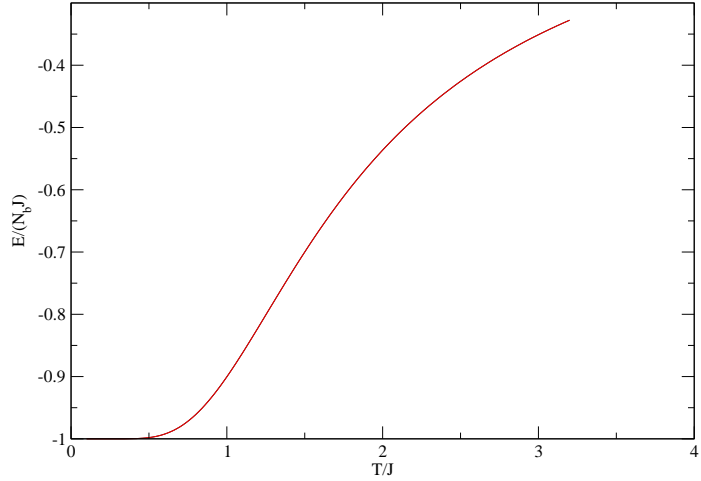}
\caption{ 
Internal energy normalized per bond and per unit coupling, $e=E/(JN_b)$, 
for the two dimensional plus-minus spin glass model. 
The calculation of $e$ is reported in Appendix B.}
\label{energy_sp_2D}
\end{figure}

For $\phi_I$ in two dimensions we can use the Onsager's solution. 
If we indicate with $z_h=\tanh(k_h)$ and $z_v=\tanh(k_v)$ 
the horizontal and the vertical parameters of the model,
Onsager's formula in the thermodynamic limit gives \cite{MCCOY}
\begin{eqnarray}\fl
\label{ONSAGER}
\varphi_I\left(z_h,z_v\right)=
\lim_{L\rightarrow \infty} \frac{1}{L} \phi_I\left(z_h,z_v\right)= 
\lim_{L\rightarrow \infty} \frac{1}{L} \log\left(\sum_{\gamma}
\prod_{b\in \gamma}z_b\right) = \nonumber \\
\frac{1}{8\pi^2} \int_{0}^{2\pi}\int_{0}^{2\pi}d\theta_h d\theta_v
\log[1+z_h^2 z_v^2+z_h^2+z_v^2+ \nonumber \\
2z_hz_v\left(z_h\cos\left(\theta_v\right)+z_v\cos
\left(\theta_h\right)\right)+ 
-2z_h\cos\left(\theta_h\right)-2z_v\cos\left(\theta_v\right)].
\end{eqnarray} 
In the case of an isotropic spin glass, we have $d\mu_h\equiv d\mu_v$, 
so that it is enough to consider the related Ising 
model with $z\equiv z_h=z_v$, and
Eq. (\ref{ONSAGER}) simplifies in 
\begin{eqnarray}
\label{ONSAGERI}
\varphi_I\left(z\right) = 
\log\left(\frac{1+z^2}{\sqrt{2}}\right)+ 
\frac{1}{\pi}\int_{0}^{\frac{\pi}{2}} 
d\omega\log [1+\left(1-\kappa^2\sin^2
\left(\omega\right)\right)^{\frac{1}{2}}], 
\end{eqnarray} 
where 
\begin{eqnarray}
\label{k}
\kappa\equiv 4z\frac{1-z^2}{\left(1+z^2\right)^2}.
\end{eqnarray} 
As is known, $\varphi_I\left(z\right)$ is non analytic at $\kappa=\pm1$, 
for which the specific heat, 
as a consequence, has a logarithmic divergence in the temperature.
In the high temperature expansion representation, 
the condition $\kappa=\pm1$ amounts to $w^{(I)}=\pm(\sqrt{2}-1)$.  
This implies that in $D=2$ the pure Ising
model presents a ferromagnetic or antiferromagnetic second order 
phase transition 
at the critical value $k_c^{(I)}=0.4407$ 
($T_c=2.269 |J|$, if the Boltzmann's constant is taken as 1).
From Eq. (\ref{logZ3}) we see therefore that our approach for 
a two dimensional Ising spin
glass would predict a second order phase transition at $\beta_c$
solution of
\begin{eqnarray}
\label{kc2}
\int d\mu \tanh^2(\beta^{(SG)} J)=\pm(\sqrt{2} -1).
\end{eqnarray}
Of course, this equation may have solutions only if the r.h.s. is positive
so that only a correspondence with a ``ferromagnetic''-like 
transition is possible. 
In particular for the plus-minus distribution (\ref{plusminus}) centered
in $J$ and $-J$, a second order phase transition would take place if 
$\tanh^2(\beta^{(SG)}_c J)=\sqrt{2}-1$,
which has solution for 
$k^{(SG)}_c=\beta^{(SG)}_c J=0.7642$ ($T^{(SG)}_c=1.308 J$).
This result is in contrast with the (by now) known fact from
numerical simulations that, in two dimensions, 
the Ising spin glass has no phase transition
at finite temperature \cite{Bhatt}. 
As is evident from subsection 9.1,
$D=2$ represents a singular case where the mapping is not definite 
since inside a bidimensional space there is only one plane.
This explains why in our approach in two dimensions 
we find a finite critical temperature; for the two dimensional
case, an effective mapping with a suitable Ising model
is impossible. The fact that there is not
a finite temperature phase transition, implies that
in two dimensions the non Ising contributions not only
are not negligible, but in the thermodynamic limit
constitute the leading part of the high temperature expansion and
hide the Ising like effects.
Nevertheless, it is interesting to see that some 
features of this model may be explored even in our Ising-like approach.
For example, as appears by comparing the plot of Fig. \ref{energy_sp_2D} for
the internal energy with the exact numerical results obtained in
\cite{Morgenstern}, though the internal energy is wrong, from its slope,
away from the critical point, we get a  
certain evaluation of the specific heat. 
Note also that, 
even though it is clearly wrong the result for the critical temperature,
according to the general rule (\ref{SGI}),
in our approximation, the effect of the disorder 
has however decreased the critical
temperature ($T^{(SG)}_c=1.308 J$) with respect to the critical temperature
of the pure Ising model ($T_c=2.269 J$).

\subsection{Three dimensional case and higher dimensions}
For $D>2$ there is no analytical solution either
for the pure Ising model and for the Ising spin glass model.
Nevertheless, several very accurate numerical (and partly
analytical) data are nowaday available for $D=3$.
From high and low temperature expansions and other numerical techniques
it is known that in $D=3$, the pure Ising model
has a critical point at $k_c^{(I)}=0.221$ ($T_c=4.51154 J$) \cite{Salman},
so that $w^{(I)}=\tanh(k_c^{(I)})=0.2180$.
Therefore, from the general rule (\ref{mapp0}), we obtain that
in $D=3$ an Ising spin glass has a 
phase transition at a critical temperature $\beta^{(SG)}_c$ such that
\begin{eqnarray}
\label{kc4}
\int d\mu \tanh^2(\beta^{(SG)}_c J)=0.2180.
\end{eqnarray}
In particular, for the plus-minus distribution,
the above equation gives 
$k^{(SG)}_c=\beta^{(SG)}_c J=0.5062$ ($T^{(SG)}_c=1.975$).
On the other hand from \cite{Rieger} a critical temperature
at $k^{(SG)}_c=0.851$ ($T^{(SG)}_c=1.175$) is found.


Similarly, in higher dimensions, 
we find more and more agreement
between our scheme and known results \cite{4D,4DG,5D},
the difference being in good accordance with our estimation
of the relative error whose order is expected to be $1/D$.

For sufficiently high, but finite dimensions, we can write a general
formula. Let us consider an hypercubic $D$ dimensional spin glass model with a 
disorder having variance $1/2D$. 
According to Eqs. (\ref{HI}-\ref{JIb}), the related Ising model
has the following Hamiltonian
\begin{eqnarray}
\label{HI1}
H_{I}=-\sum_{b\in\Gamma} J \tilde{\sigma}_b,
\end{eqnarray}
so that, if  $D$ is large enough it has a critical temperature located a 
\cite{Fisher}
\begin{eqnarray}
\label{betaI0}
\beta_c^{(I)} J 2D=1 - \frac{1}{2D} + O\left(\frac{1}{D^2}\right).
\end{eqnarray}
Therefore, by expanding at the first order in $1/D$ Eq. (\ref{mapp0}),
we find that the critical temperature for the corresponding 
spin glass is given by
\begin{eqnarray}
\label{betaSG}
\beta_c^{(SG)} =1 +O\left(\frac{1}{D}\right),
\end{eqnarray}
in accordance with the result of references \cite{Fisher1} and  
\cite{Mezard} in which a perturbative expansion in
powers of $1/D$ is performed.


\section{Applications to infinite dimensional models}
As $D\rightarrow \infty$ our scheme becomes exact. 
In particular, for the SK model, above the critical 
temperature it reproduces the mean field solution, 
even though the procedure is completely
different from the replica and cavity methods \cite{Parisi}. 
Notice that, as we have just anticipated in the Sec. 4, 
our approach, despite of the fact that in $D=\infty$ 
becomes exact, gives access to the free energy only above 
the critical temperature and,
an analytic continuation below the critical temperature is not
allowed. Note, in particular, that a forced use of the mapping to
calculate the free energy at low temperature, would give a completely
wrong result. In fact, given a system in $D$ dimensions, it is easy
to see that $\varphi_I(\beta)/\beta\to 0$ for $\beta\to\infty$, so that,
at zero temperature, from Eq. (\ref{logZ2}) applied to
a plus-minus measure, it remains only $F_0/N=U_0/N=-JD$, 
whereas the Derrida lower bound
for large $D$ gives $U_0/N\sim -J\sqrt{2D\log(2)}$ \cite{Derrida}.
However, as just pointed out in the subsection 4.3, we argue that an
analytic continuation of other physical quantities, such as
the crossover surfaces or the magnetizations, provide
a certain effective approximation. 

\subsection{Sherrington Kirkpatrick model}
The SK model corresponds to the spin glass over the fully connected graph
and anyone of the $N$ spins 
interacts with any other spin trough a random coupling $J_b$ whose probability
distribution has homogeneous rescaled mean value and variance 
given respectively by
\begin{eqnarray}
\label{dmugauss1}
\mediau{J_b} = J_0/N, \\
\label{dmugauss2}
\mediau{(J_b-J_0/N)^2}=\tilde{J}^2/N.
\end{eqnarray} 

\begin{figure}[t]
\centering
\includegraphics[width=0.5\columnwidth,clip]{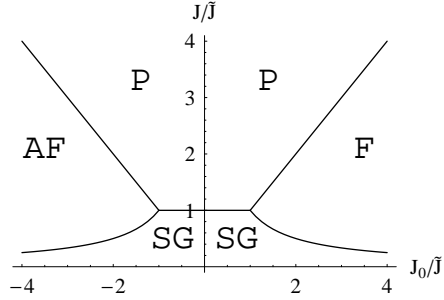}
\caption{ 
Phase diagram of the Sherrington Kirkpatrick model. 
The upper line, \textit{i.e.} the envelope of the $P-SG$, 
the $P-F$ and the $P-AF$ lines, obtained from 
Eqs. (\ref{MF4}) and (\ref{MF5}), is exact. The crossover curves 
$SG-F$ and $SG-AF$ are obtained from Eq. (\ref{MF6}).}
\label{SK}
\end{figure}

From Eqs. (\ref{JI}) we see that
the Hamiltonian of the related Ising model is 
\begin{eqnarray}
\label{MF}
H_{I}=-\sum_{b\in \Gamma_f} J^{(I)} \tilde{\sigma}_b=
-\sum_{(i,j)} J^{(I)} \sigma_i\sigma_j,
\end{eqnarray} 
where in the last expression we have rewritten $H_{I}$ in the usual form 
as a sum over the couples of sites $(i,j)$. 
As is well known, for this model, depending on the sign of the 
coupling $J^{(I)}$, a ferromagnetic-paramagnetic or
an antiferromagnetic-paramagnetic phase transition takes
place at the same critical temperature given by \cite{Trizac}
\begin{eqnarray}
\label{betaI}
\beta_c^{(I)}|J^{(I)}|N=1,
\end{eqnarray} 
which for $N$ large, in terms of the universal quantities 
$w_{F/AF}^{(I)}=\pm\tanh(\beta_c^{(I)}|J^{(I)}|)$ gives
\begin{eqnarray}
\label{betaI1}
w_{F/AF}^{(I)}=\pm\frac{1}{N}+O\left(\frac{1}{N^3}\right)
\end{eqnarray} 
On the other hand by using Eqs. (\ref{dmugauss1}-\ref{dmugauss2}) 
for $N$ large we have
\begin{eqnarray}
\label{MF2}
\mediau{\tanh^2\left(\beta J_b\right)}=
\frac{\left(\beta\tilde{J}\right)^2}{N}+O\left(\frac{1}{N^3}\right),  
\end{eqnarray}
and
\begin{eqnarray}
\label{MF3}
\mediau{\tanh\left(\beta J_b\right)}=
\frac{\left(\beta J_0\right)}{N}+O\left(\frac{1}{N^3}\right).
\end{eqnarray}
Therefore, from Eqs. (\ref{mapp0}) and (\ref{mapp01}), in the
limit $N\to \infty$, we get
the following spin glass $(SG)$ and, depending on the sign of 
$J_0$, ferromagnetic $(F)$ or antiferromagnetic $(AF)$ phase boundaries 
\begin{eqnarray}
\label{MF4}
\beta_c^{(SG)}\tilde{J}&=& 1 \\
\label{MF5}
\beta_c^{(F/AF)}J_0&=&\pm 1.
\end{eqnarray}
Finally, according to Eq. (\ref{mapp}), by taking the envelope
of the curves (\ref{MF4}) and (\ref{MF5}) we get the upper
phase boundaries shown in Fig. \ref{SK}.
In the same figure we report also the coexistence curves $SG-F$ and $SG-AF$
derived from the systems (\ref{coexF}-\ref{coexAF}) whose solution is given by
\begin{eqnarray}
\label{MF6}
\left(\beta\tilde{J}\right)^2=\beta |J_0|.
\end{eqnarray}
Note however that, unlike the upper
phase boundaries, the coexistence curves are not exact;
they are only representative of the true coexistence curves.

In infinite dimensions, 
our approach becomes exact only above the upper phase boundaries,
where the high temperature part of the free energy, 
$\varphi$, becomes trivially 0.
On the other hand, an analytic continuation below these
boundaries is not allowed. This fact can be understood
considering that the thermodynamic of the related Ising
model (\ref{MF}) turns out to be well defined only above the
critical temperature, which in the thermodynamic limit
is infinite ($\beta_c^{(I)}\to 0$); 
for a well defined thermodynamic the $J^{(I)}$
in the Hamiltonian (\ref{MF}) should be replaced by $J^{(I)}/N$.
Nevertheless, via analytic continuation,
besides the extrapolation for the coexistence curves (\ref{MF6}),
a simple estimation of the Edward Anderson parameters and of the
magnetizations is possible. In Appendix C we show that 
in the SK model, the square root of Edward Anderson parameter 
and the magnetization, respectively indicated as 
$m_{SG}$ and $m_{F}$, naturally emerge as effective fields in
correspondence with the fields of the related Ising model.
For $J_0\geq 0$, they are 
solution of the following mean field equations, respectively
\begin{eqnarray}
\label{MF10}
m_{SG}=\tanh\left((\beta\tilde{J})^2m_{SG}\right),
\end{eqnarray}
and 
\begin{eqnarray}
\label{MF11}
m_{F}=\tanh\left(\beta J_0m_{F}\right).
\end{eqnarray}
Note however, that the above expressions are not exact.
In particular, the slope of $m_{SG}$ near the critical
point is wrong.

Similarly, when $J_0<0$ the square root of the Edward Anderson parameter 
and the magnetization, which now are described by two 
effective fields, $m^{(a)}$ and $m^{(b)}$ related to
the two sublattices $a$ and $b$ in which the initial lattice
can be decomposed, are solutions of 
the following mean field systems, respectively
\begin{eqnarray}
\label{MF17} 
\left\{
\begin{array}{c}
m^{(a)}_{SG}= -\tanh\left((\beta\tilde{J})^2m_{SG}^{(b)}\right), \\
m^{(b)}_{SG}= -\tanh\left((\beta\tilde{J})^2m_{SG}^{(a)}\right),
\end{array}
\right.
\end{eqnarray}
and
\begin{eqnarray}
\label{MF18} 
\left\{
\begin{array}{c}
m^{(a)}_{AF}= -\tanh\left(\beta J_0 m_{AF}^{(b)}\right), \\
m^{(b)}_{AF}= -\tanh\left(\beta J_0 m_{AF}^{(a)}\right).
\end{array}
\right.
\end{eqnarray}

\subsection{SK generalized; random antiferromagnets}
One interesting generalization of the SK model,
is a model defined over a lattice which
can be decomposed in many, say $p$, sublattices 
eachone constituted of $N$ sites and with the
constrain that a spin over one sublattice interacts
only with the spins over the other sublattices.
Such systems are particularly interesting as
models of random antiferromagnets. 
The Hamiltonian 
is given by \cite{Almeida}
\begin{eqnarray}
\label{MF19}
H=-\sum_{(\mu,\nu)}\sum_{(i,j)}J_{(i,j)}^{\mu,\nu}
\sigma_{i,\mu}\sigma_{j,\nu},
\end{eqnarray}
where the latin indeces $i,j=1,\ldots,N$, label the sites 
in each sublattice, whereas the greek indeces $\mu,\nu,=1,\ldots,p$, 
label the sublattices. 
A coupling $J_{(i,j)}^{\mu,\nu}$ is then in correspondence with the bond
$b=(i,\mu;j,\nu)$ connecting the site $i$ of the sublattice
$\mu$ with the site $j$ of the sublattice $\nu$.
Let us consider for simplicity a distribution with a homogeneous rescaled 
mean value and variance given respectively by
\begin{eqnarray}
\label{MF20}
\mediau{J_{(i,j)}^{\mu,\nu}} &=& \frac{J_0}{N(p-1)}, \\
\label{MF21}
\mediau{\left(J_{(i,j)}^{\mu,\nu}-\frac{J_0}{N(p-1)}\right)^2}&=&
\frac{\tilde{J}^2}{N(p-1)}.
\end{eqnarray} 
The related Ising model has Hamiltonian
\begin{eqnarray}
\label{MF22}
H_I=-\sum_{(\mu,\nu)}\sum_{(i,j)}J^{(I)}
\sigma_{i,\mu}\sigma_{j,\nu},
\end{eqnarray}
Following \cite{Anderson}, the most general solution of
the related Ising model must be found
by introducing $p$ effective fields $m_I^{(1)},\ldots,m_I^{(p)}$
satisfying the system of equations 
\begin{eqnarray}
\label{MF23}
m_I^{(l)}=-\sum_{k\neq l}\tanh\left(-\beta m_I^{(k)}J^{(I)}N\right),
\quad l=1,\ldots,p.
\end{eqnarray}
Note that the sign in front
of $H_I$ is, for convenience, reversed with respect to \cite{Anderson};
we recall that the coupling $J^{(I)}$ of the related Ising
model is homogeneous, but its value can be arbitrary.
Note also that, with respect to the convention adopted in \cite{Almeida},
the sign for the parameter $J_0$ has been reversed.

\begin{figure}[t]
\centering
\includegraphics[width=0.5\columnwidth,clip]{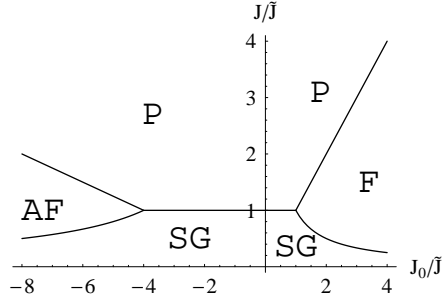}
\caption{ 
Phase diagram of the generalized Sherrington Kirkpatrick model 
(\ref{MF19}) with $p=5$. The diagram can be compared with that
reported in \cite{Almeida}.
The upper line, \textit{i.e.} the envelope of the $P-SG$, 
the $P-F$ and the $P-AF$ lines, obtained from 
Eqs. (\ref{MF29}-\ref{MF31}), is exact. The crossover curves 
$SG-F$ and $SG-AF$ are obtained from Eqs. (\ref{MF32}) and (\ref{MF33}).}
\label{AF}
\end{figure}

Linearizing Eq. (\ref{MF23}) for small fields we arrive 
at the homogenous system
\begin{eqnarray}
\label{MF24}
m_I^{(l)}=-x\sum_{k\neq l}m_I^{(k)}, \quad l=1,\ldots,p
\end{eqnarray}
where $x$ is given by
\begin{eqnarray}
\label{MF25}
x=-\beta J^{(I)}N.
\end{eqnarray}
The homogenous system (\ref{MF24}) has a non zero solution
for values of $x$ for which the matrix of the coefficients $A$ has
determinant zero. It is easy to see that this determinant is
simple given by 
\begin{eqnarray}
\label{MF26}
\det(A)=\left(1-x\right)^{p-1}\left(1+(p-1)x\right),
\end{eqnarray}
therefore, the system will have a non zero solution for
the critical values $x=1$ or $x=-1/(p-1)$, which using Eq. (\ref{MF25}) means
\begin{eqnarray}
\label{MF27}
-\beta_c^{(I)} J^{(I)}N=\left\{
\begin{array}{c}
\frac{-1}{p-1}, \\
1
\end{array}
\right.
\end{eqnarray}
Recalling the explicit sign in front of the Hamiltonian (\ref{MF22}),
we note that the first solution corresponds to an ordinary 
Ising ferromagnet with $J^{(I)}\geq 0$, whereas the second one corresponds 
to a generalized antiferromagnetic Ising model with $J^{(I)}<0$
and with $p$ sublattices.
For $N$ large, in terms of the universal quantities 
$\tanh(\beta_c^{(I)} J^{(I)})$, Eq. (\ref{MF27}) gives
\begin{eqnarray}
\label{MF28}
w_{F}^{(I)}&=& \frac{1}{N(p-1)} \\
\label{MF29}
w_{AF}^{(I)}&=& \frac{-1}{N},
\end{eqnarray}
which, according to the general rule (\ref{mapp0}-\ref{mapp01}) 
and by using Eqs. (\ref{MF20}-\ref{MF21}) bring to the 
following spin glass $(SG)$ and, depending on the sign of 
$J_0$, ferromagnetic $(F)$ or antiferromagnetic $(AF)$, phase boundaries 
\begin{eqnarray}
\label{MF29}
\beta_c^{(SG)}\tilde{J}&=&1, \\
\label{MF30}
\beta_c^{(F)}J_0&=& 1, \quad J_0\geq 0,\\
\label{MF31}
\beta_c^{(AF)}J_0&=& -(p-1), \quad J_0<0
\end{eqnarray}
Finally, according to Eq. (\ref{mapp}), by taking the envelope
of the curves (\ref{MF29}-\ref{MF31}) we get the upper
phase boundaries shown in Fig. \ref{AF}.
In the same figure we report also the coexistence curves $SG-F$ and $SG-AF$
derived from Eqs. (\ref{coexF}-\ref{coexAF}) whose equations are given by
\begin{eqnarray}
\label{MF32}
\left(\beta\tilde{J}\right)^2=\beta J_0, \quad J_0\geq 0,
\end{eqnarray}
for the $SG-F$ crossover, and
\begin{eqnarray}
\label{MF33}
\left(\beta\tilde{J}\right)^2=\frac{-\beta J_0}{p-1}, \quad J_0<0,
\end{eqnarray}
for the $SG-AF$ crossover. 

Note that for $p=2$ one recovers the standard SK model. 

\subsection{Spin glass over Bethe lattice}
Let us consider a spin glass model defined over a Bethe lattice of
coordination number $q$ 
(\textit{i.e.}, branching number $k=q-1$, see definition in Sec. \ref{GRAPH}).
In the case of a homogeneous model, $d\mu_b\equiv d\mu$, 
the related Ising model is the 
homogeneous Ising model over a regular Bethe lattice with coordination 
number $k$ for which the exact solution is known \cite{Baxter}.
This solution predicts a second order phase transition at a value
of $\beta_c^{(I)}$ given by
\begin{eqnarray}
\label{MF34}
\beta_c^{(I)}J^{(I)}=\frac{1}{2}\log\left(\frac{q}{q-2}\right).
\end{eqnarray}
In terms of the universal quantity $w_F^{(I)}=\tanh(\beta_c^{(I)} J^{(I)})$,
Eq. (\ref{MF34}) reads
\begin{eqnarray}
\label{MF35}
w_F^{(I)}=\tanh\left(\beta_c^{(I)}J^{(I)}\right)=\frac{1}{q-1},
\end{eqnarray}
where we have made use of the hyperbolic identity 
\begin{eqnarray}
\label{MF36}
\tanh\left(\frac{1}{2}\log(r) \right)=\frac{r-1}{r+1}.
\end{eqnarray}
Therefore, by using the general rule (\ref{mapp0}-\ref{mapp01})
we find that a spin glass ($SG$) and a disordered ferromagnetic ($F$)
transitions take place at $\beta_c^{(SG)}$ and $\beta_c^{(F)}$, respectively 
solutions of the two following equations
\begin{eqnarray}
\label{MF37}
\mediau{\tanh^2\left(\beta_c^{(SG)}J_b\right)}&=&\frac{1}{q-1}, \\
\label{MF38}
\mediau{\tanh\left(\beta_c^{(F)}J_b\right)}&=&\frac{1}{q-1}.
\end{eqnarray}
Finally, by using Eq. (\ref{mapp}) and Eqs. (\ref{coexF}-\ref{coexAF}),
the upper phase boundary and the crossover surfaces follow.

\section{Conclusions}
In the framework of the high temperature expansion,
we have derived a general mapping between an Ising spin glass model
and a related Ising one, 
Eqs. (\ref{mapp0}-\ref{mapp3}) and (\ref{mapp0g}-\ref{mapp3g}).
The mapping is definite above the critical temperature and
becomes exat in all the paramagnetic region when
the dimension $D\to\infty$ in the strict sense,
whereas, more in general, becomes exact in the limit
$\beta\to\beta_c^{-}$ when the dimension $D\to\infty$
in the large sense, where now $D$, roughly speaking, 
is the number of independent paths per site (see Sec. 3).
The mapping can be applied to find exactly the upper phase boundaries
of a spin glass model if the critical temperature of the related
Ising model is known.
Furthermore, by analytic continutation, the mapping provides even 
information such as the crossover surfaces, the magnetizations,
and the correlation functions. Even if this further
application is not exact, we argue that its use can turn out to
be quite effective as a first insight to the physics of very complex models.

We have applied the mapping to several known models at finite and infinite 
dimensions. In particular, the last application 
of a spin glass model defined over a Bethe lattice suggests a comment. 
We observe, in fact, that the relations for the Bethe lattice, 
Eqs. (\ref{MF37}-\ref{MF38}), 
have been derived by many authors in several contexts and by using different 
procedures, see for example \cite{Viana}-\cite{MezardP}.
However, it is quite impressive the easiness by which we have derived 
these formulas (and similarly those for the SK models);
simply starting from the solution of the related
Ising model. For example, within the limits that our approach
concerns only the singularities of the free energy,
we had not to be worried about
the rigorous definition of the Bethe lattice. 
As is known, in fact, the so called Baxter exact solution
of the Bethe lattice concerns, more precisely, the inner
part of an infinite Cayley tree far, in a sense, from the boundary
in which a finite fraction of
the total number of spins resides (see Sec. \ref{GRAPH}). If one does not
exclude in some way these boundaries, the model turns out to
be greatly sensitive to the boundary conditions and difficult
to treat. 
Another way to avoid this drawback consists in considering for example
an ensemble of uncorrelated random graphs in which any
site has a mean connectivity equal to $q$.
We stress however that, as far as one considers only
the singularities of the free energy, whatever the 
lattice and the system of links $\Gamma$
over which we build a spin glass model could be complicated,
in our approach, all the mathematical and technical difficulties, 
even those connected to the delicate questions concerning 
boundary conditions, if any,  
are completely reduced at the level of the corresponding
related Ising model, which, as a non random
model, is remarkable very simpler to solve. 
It is also clear that this difference becomes particularly important if the
given model has many degrees of freedom (mechanical or disorder parameters). 
In our approach, once the phase boundary of 
an Ising model in infinite dimension 
is known, this solution can be immediately applied to
find the upper critical boundary of 
the corresponding spin glass model for which
the Ising one turns out to be the related Ising model
according to the definition given in Sec. 4.
Hence, for example, since the Ising model over a Bethe lattice
dos not depend on the chosen boundary conditions, the upper
critical surface of an Ising spin glass model over the Bethe lattice
will be the same for any chosen boundary condition as well,
regardeless of the fact that the non paramagnetic regions
can be different (see comment in Sec. 11).
 
We want to poin out also the numerical advantage that our mapping implies.
In fact, even if the analytical knowledge of the critical temperature of
some Ising model in not known, an its numerical evaluation turns
out to be hugely easier than a direct numerical evaluation for
the corresponding Ising spin glass model. Once a numerical estimation
for the critical temperature $1/\beta_c^{(I)}$ is known, 
the mapping returns immediatly the
upper critical surface of the spin glass.

The generalization of the mapping to include also a randomness
of the set of links, necessary to study random models on 
random graphs (see, \textit{e.g.}, \cite{Dorogovstev}), 
will be presented in a forthcoming work (part II). 
Finally, we observe that the simple argument we have used
to prove our mapping, ``random Ising system '' $\rightarrow$ 
``non random Ising system'', which is exact in infinite dimensions, 
does not seem to be peculiar of the Ising model, 
so that, even for more general models, 
a mapping at high temperature 
``random system'' $\rightarrow$ ``non random system'' appears possible.

\section*{Acknowledgments}
This work was supported by DYSONET under NEST/Pathfinder initiative FP6, 
and by the FCT (Portugal) grant SFRH/BPD/24214/2005. 
The research was also partially supported by Italian MIUR under 
PRIN 2004028108$\_$001. 
I am grateful to A. V. Goltsev for a critical reading of the manuscript
and useful discussions.

\appendix

\addcontentsline{toc}{section}{Free energy in the presence of external fields}
\section*{Appendix A. Free energy in the presence of external fields}
\setcounter{section}{1}

The result of Sec. 9 can be formally generalized  
for arbitrarly external fields $\{h_i\}$.
In this case, Eq. (\ref{Z2}) is to be modified as follows
\begin{eqnarray}\fl
\label{Z2h}
Z\left(\{J_b\};\{h_i\}\right) &=& 
\prod_{b\in\Gamma} \cosh\left(K_b\right) \prod_{i=1}^N \cosh\left(H_i\right)
\nonumber \\ && \times \sum_{\{\sigma_i\}}
\prod_{b\in\Gamma} \left(1+\tilde{\sigma}_b\tanh\left(K_b\right)\right)
\prod_{i=1}^N \left(1+\sigma_i\tanh\left(H_i\right)\right),
\end{eqnarray} 
where we have introduced the symbol
\begin{eqnarray}
\label{Hh}
H_i=\beta h_i.
\end{eqnarray} 
Correspondingly, the high temperature expansion of $Z$
will be a series over arbitrary combinations 
of closed and open paths. Given one of these generalized paths, 
the weight of a bond $b$, as before, is provided by $\tanh(K_b)$, 
whereas the weight of two extremal points $i$ and $j$ 
of an open path are provided by 
$\tanh(H_i)$ and $\tanh(H_j)$, respectively.
Taking into account that the factors $\tanh(H_i)$'s are not affected
by the integration over the $J_b$'s, the generalization
of Eq. (\ref{PAn2}) to generic external field follows
straightforward, where, along with the bond-weights $F_b^{(p)}$,
now appear also the point-weights $\tanh^q(H_i)$,
where $p$ and $q$ count, respectively, the number of overlapping bonds
and points, among the $n$ replica generalized paths.
On the other hand, it is not difficult to see that 
we can apply the argument on the dimensionality
even to open paths and to combinations of these with
closed paths so that even in the presence of 
an external field the mapping of Eqs. (\ref{mapp1}-\ref{mapp3}) follows.
The problem here is that this argument is only formal. In fact,
if a nonzero magnetic field is present, 
with respect to the parameters $z_b=\tanh(K_b)$, the high temperature
expansion in general can have a zero radius of convergence when $D=\infty$, 
and this is the case for the related Ising model; for $h_i\neq 0$
one has in general a non zero mean magnetization, so that the density energy
of the related Ising model goes to infinite as $D\to \infty $.

\addcontentsline{toc}{section}{Internal energy for the $D=2$ spin glass}
\section*{Appendix B. Internal energy for the $D=2$ spin glass}
\setcounter{section}{2}

In this appendix we evaluate the free energy and the internal
energy for the two dimensional spin glass with a plus-minus measure.
As stressed in Sec. (\ref{Two}) we get only a rough approximation,
but it provides however a case in which an analytical calculation
is possible.  

According to the general
rule of Eqs. (\ref{mapp1}-\ref{mapp3}) which, specifically,
for a centered measure amount to Eqs. (\ref{logZ3}-\ref{Fb2}),
the free energy is readily obtained 
simply by using for $\varphi$ Eqs. (\ref{ONSAGERI}) and (\ref{k})
with the mapping
\begin{eqnarray}
\label{A1}
\varphi_I\left(\tanh(\beta J)\right)\rightarrow 
\varphi=\frac{1}{2}\varphi_I\left(\tanh^2(\beta J)\right).
\end{eqnarray}
The internal energy $U$ can be derived from 
the free energy $F$ by using the formula
\begin{eqnarray}
\label{A2}
U=\frac{\partial{(\beta F)}}{\partial \beta}.
\end{eqnarray}
Therefore, from Eq. (\ref{logZ3}) for the free energy per bond and
per unit of coupling, $e=E/(J N_b)$, we have 
\begin{eqnarray}
\label{A3}
e=-\tanh(\beta J) - \frac{1}{2J}\frac{\partial{\varphi}}{\partial \beta},
\end{eqnarray}
where the factor $2$ in the denominator of the second term takes into
account that for any two sites we have a bond ($N_b=N/2$).
The derivative of $\varphi$ can be calculated as follows
\begin{eqnarray}
\label{A4}
\frac{\partial{\varphi}}{\partial \beta}=\frac{\partial{\varphi}}{\partial z}
\frac{\partial{z}}{\partial \beta},
\end{eqnarray}
\begin{eqnarray}\fl
\label{A5}
\frac{\partial{\varphi}}{\partial z}=\frac{2z}{1+z^2}
-\frac{1}{\pi}\int_{0}^{\frac{\pi}{2}}\kappa\frac{\partial \kappa}{\partial z} 
d\omega\frac{\sin^2\left(\omega\right)}
{1+\left(1-\kappa^2\sin^2\left(\omega\right)\right)^{\frac{1}{2}}}
\frac{1}{\left(1-\kappa^2\sin^2\left(\omega\right)\right)^{\frac{1}{2}}}, 
\end{eqnarray}
\begin{eqnarray}
\label{A6}
\frac{\partial{z}}{\partial \beta}=2J \frac{\tanh(\beta J)}{\cosh^2(\beta J)},
\end{eqnarray}  
whereas $\kappa$ is given by Eq. (\ref{k}) and its derivative is
\begin{eqnarray}
\label{A7}
\frac{\partial{\kappa}}{\partial z}=4\frac{1-z^2}{(1+z^2)^2}-
8\frac{z^2}{(1+z^2)^2}-164z^2\frac{1-z^2}{(1+z^2)^3}.
\end{eqnarray}
By inserting the expressions (\ref{A4}-\ref{A6}) in Eq. (\ref{A3})
and by using for $\varphi$ the mapping (\ref{A1}), 
one gets the free internal energy $e$ whose plot is reported
in Fig. (\ref{energy_sp_2D}).

\addcontentsline{toc}{section}{Effective fields}
\section*{Appendix C. Effective fields}
\setcounter{section}{3}

In our approach the key ingredient is the knowledge 
of the non trivial part of the
free energy $\phi_I$ of the related Ising model, in terms of which
the free energy is written as
\begin{eqnarray}\fl
\label{BMF1}
-\beta F_I &=& N\log(2)+\frac{N(N-1)}{2} 
\log\left(\cosh\left(\beta J^{(I)} \right)\right) 
+ \phi_I\left(\tanh\left(\beta J^{(I)}\right)\right).
\end{eqnarray}
Once $\phi_I$ is known, according to Eqs. (\ref{mapp2}-\ref{mapp3}),
in order to calculate $\phi$, 
we have to implement in $\phi_I$ the following transformations
\begin{eqnarray}
\label{BMF2}
\tanh\left(\beta J^{(I)}\right) \rightarrow 
\mediau{\tanh^2\left(\beta J_b\right)}=
\frac{\left(\beta\tilde{J}\right)^2}{N}+O\left(\frac{1}{N^3}\right),  
\end{eqnarray}
and
\begin{eqnarray}
\label{BMF3}
\tanh\left(\beta J^{(I)}\right )\rightarrow 
\mediau{\tanh\left(\beta J_b\right)}=
\frac{\left(\beta J_0\right)}{N}+O\left(\frac{1}{N^3}\right).
\end{eqnarray}

As is well known, up to terms negligible in the thermodynamic limit, 
the fully connected Ising model with homogeneous coupling $J^{(I)}$, 
has a mean field solution which, depending on the sign of $J^{(I)}$
can be ferromagnetic or antiferromagnetic. 
It is convenient to distinguish the two sub-cases.

Let us suppose that $J^{(I)}\geq 0$.
In this case the mean field
solution for the related Ising model is given by
\begin{eqnarray}
\label{BMF4}
-\beta F_I = N \log\left(2\cosh\left(\beta J^{(I)} m_I N\right)\right) 
-\frac{N^2\beta J^{(I)} m_I^2}{2},
\end{eqnarray}
where $m_I$ satisfies the mean field equation
\begin{eqnarray}
\label{BMF5}
m_I= \tanh\left(\beta m_I J^{(I)} N\right).
\end{eqnarray}
From Eqs. (\ref{BMF1}) and (\ref{BMF4}-\ref{BMF5}), for the density 
$\varphi_I=\phi_I/N$ we then get 
\begin{eqnarray}
\label{BMF6}\fl
\varphi_I \left(\tanh\left(\beta J^{(I)}\right)\right)&=& 
\log\left(\cosh\left(\beta J^{(I)} m_I N\right)\right)
-\frac{N}{2}\log\left(\cosh\left(\beta J^{(I)} \right)\right) \nonumber \\
&& \left. -\frac{\beta J^{(I)} m_I^2 N}{2}
\right|_{m_I= \tanh\left(\beta m_I J^{(I)}N\right)}
\end{eqnarray}

Equation (\ref{BMF6}) must now be used in the 
mapping of Eqs. (\ref{mapp2}-\ref{mapp3}), by using
the transformations (\ref{BMF2}-\ref{BMF3}).
To this aim it is convenient to use the relation 
$\cosh^2(x)=1/(1-\tanh^2(x))$ and to rewrite Eq. (\ref{BMF6}) as 
\begin{eqnarray}
\label{BMF7}\fl
\varphi_I \left(\tanh\left(\beta J^{(I)}\right)\right)= 
-\frac{1}{2}\log\left( 1-m_I^2 \right)
+\frac{N}{4}\log\left(1-\tanh^2\left(\beta J^{(I)} \right)\right) 
\nonumber \\ \left. 
-\frac{ \arctanh\left(\tanh\left(\beta J^{(I)} \right)\right) m_I^2 N}{2}
\right|_{m_I= 
\tanh\left(\arctanh\left(\tanh\left(\beta J^{(I)}\right)\right)m_I N\right)}
\end{eqnarray}
By inserting the transformations (\ref{BMF2}) and (\ref{BMF3}) 
in Eq. (\ref{BMF7}) for $N$ large we arrive, respectively, at
\begin{eqnarray}
\label{BMF8}
\varphi_I \left(\mediau{\tanh^2\left(\beta J\right)}\right)= 
-\frac{1}{2}\log\left( 1-m_{SG}^2 \right) 
-\frac{\left(\beta \tilde{J}\right)^2 m_{SG}^2}{2},
\end{eqnarray}
and
\begin{eqnarray}
\label{BMF9}
\varphi_I \left(\mediau{\tanh\left(\beta J\right)}\right)= 
-\frac{1}{2}\log\left( 1-m_{F}^2 \right)  
-\frac{\beta J_0m_{F}^2}{2},
\end{eqnarray}
where the effective fields $m_{SG}$ and $m_{F}$ have been obtained,
respectively, by using the transformations (\ref{BMF2}) and (\ref{BMF3})
and are given by
\begin{eqnarray}
\label{BMF10}
m_{SG}=\tanh\left((\beta\tilde{J})^2m_{SG}\right),
\end{eqnarray}
and 
\begin{eqnarray}
\label{BMF11}
m_{F}=\tanh\left(\beta J_0m_{F}\right).
\end{eqnarray}
The obvious interpretation of these fields is that,
whereas $m_{F}$ is the magnetization of the system (in presence
of a disorder with variance $\tilde{J}$), $m_{SG}$ represents
$\sqrt{q_{EA}}$, \textit{i.e.}, up to a square root, it is
the Edward-Anderson order parameter, which below the critical
temperature is non zero even for $J_0=0$.

Let us now suppose $J^{(I)}<0$.
In this case the mean field
solution for the related Ising model is given in terms of two
effective fields $m^{(a)}_I$ and $m^{(b)}_I$ related
to two sublattices $a$ and $b$ in which the given lattice $\Lambda$
can be decomposed. The sublattices $a$ and $b$ are defined symmetrically
so that the first neighbors of a site of the sublattice $a$ are 
sites of the sublattice $b$ and vice versa.
In terms of the effective fields $m^{(a)}_I$ and $m^{(b)}_I$
the free energy of the related Ising model is given by
\begin{eqnarray}
\label{BMF12}
-\beta F_I &=& \frac{N}{2}\log\left(
2\cosh\left(\beta J^{(I)} m^{(a)}_I N\right)
2\cosh\left(\beta J^{(I)} m^{(b)}_I N\right)
\right) \nonumber \\
&& +\frac{N^2\beta J^{(I)} m^{(a)}_Im^{(b)}_I}{2},
\end{eqnarray}
where $m^{(a)}_I$ and $m^{(b)}_I$ satisfy the mean field system
\begin{eqnarray}
\label{BMF13} 
\left\{
\begin{array}{c}
m^{(a)}_I= -\tanh\left(\beta m^{(b)}_I J^{(I)} N\right), \\
m^{(b)}_I= -\tanh\left(\beta m^{(a)}_I J^{(I)} N\right).
\end{array}
\right.
\end{eqnarray}
Similarly to what done in the previous case, from Eqs. (\ref{BMF1}) and
(\ref{BMF12}-\ref{BMF13}), for the density $\varphi_I$ we get
\begin{eqnarray}\fl
\label{BMF14}
\varphi_I \left(\tanh\left(\beta J^{(I)}\right)\right)&=&
\frac{1}{2}\log\left(
\cosh\left(\beta J^{(I)} m^{(a)}_I N\right)
\cosh\left(\beta J^{(I)} m^{(b)}_I N\right)
\right) \nonumber \\
&& +\frac{N\beta J^{(I)} m^{(a)}_Im^{(b)}_I}{2}
-\frac{N}{2}\log\left(\cosh\left(\beta J^{(I)} \right)\right), 
\end{eqnarray}
and by inserting the transformations (\ref{BMF2}) and (\ref{BMF3}) 
for $N$ large we arrive, respectively, at
\begin{eqnarray}\fl
\label{BMF15}
\varphi_I \left(\mediau{\tanh^2\left(\beta J\right)}\right)&=& 
-\frac{1}{4}\log\left( 
\left(1-{m_{SG}^{(a)}}^2\right)
\left(1-{m_{SG}^{(b)}}^2\right) \right) \nonumber \\
&& +\frac{\left(\beta \tilde{J}\right)^2 m_{SG}^{(a)}m_{SG}^{(b)}}{2},
\end{eqnarray}
and
\begin{eqnarray}\fl
\label{BMF16}
\varphi_I \left(\mediau{\tanh\left(\beta J\right)}\right)&=& 
-\frac{1}{4}\log\left( 
\left(1-{m_{AF}^{(a)}}^2\right)
\left(1-{m_{AF}^{(b)}}^2\right) \right) \nonumber \\
&& +\frac{\beta J_0 m_{AF}^{(a)}m_{AF}^{(b)}}{2},
\end{eqnarray}
where the effective fields $m_{SG}^{(a)}$ and $m_{SG}^{(b)}$,
and $m_{F}^{(a)}$ and $m_{F}^{(b)}$, satisfy
the following mean field systems, respectively
\begin{eqnarray}
\label{BMF17} 
\left\{
\begin{array}{c}
m^{(a)}_{SG}= -\tanh\left((\beta\tilde{J})^2m_{SG}^{(b)}\right), \\
m^{(b)}_{SG}= -\tanh\left((\beta\tilde{J})^2m_{SG}^{(a)}\right),
\end{array}
\right.
\end{eqnarray}
and
\begin{eqnarray}
\label{BMF18} 
\left\{
\begin{array}{c}
m^{(a)}_{AF}= -\tanh\left(\beta J_0 m_{AF}^{(b)}\right), \\
m^{(b)}_{AF}= -\tanh\left(\beta J_0 m_{AF}^{(a)}\right).
\end{array}
\right.
\end{eqnarray}



\begin{thebibliography}{99}
\bibitem{NOTA0} Up to the question of the analytic continuation 
$n$ integer $\to n$ real, the validity of the replica symmetric solution above
the critical temperature in the (usual)
replica approach, was proved long ago in \cite{HEMMEM},
but limited to the SK model. Finally, the definitive rigorous
works of F. Guerra \cite{Guerra} and M. Talagrand \cite{Talagrand}
concern the SK and the p-spin model. Some rigorous results have been
provided even for the so called Kac models \cite{Franz}
\bibitem{Palmer2} Such an observation was made long ago in \cite{Palmer}
for the SK model
\end{thebibliography}
\end{document}